\documentclass[aip,jcp,preprint,groupedaddress,onecolumn]{revtex4-1}
\usepackage{graphicx}

\usepackage{siunitx}
\usepackage{amsmath}
\usepackage{amstext}
\usepackage{amsfonts}
\usepackage{amssymb}
\usepackage{dcolumn}
\usepackage{setspace}
\usepackage{hyperref}   
\usepackage{mdwlist}
\newcommand{\bs}[1]{{{{\bf #1}}}}
\newcommand{\Fig}[1]{Fig.\,\ref{#1}}
\newcommand{\Eq}[1]{Eq.\,(\ref{#1})}

\newcommand{\Ref}[1]{Ref.~\onlinecite{#1}}
\newcommand{\Refs}[1]{Refs.~\onlinecite{#1}}
\newcommand{\p}{^{\prime}}
\newcommand{\iu}{{i\mkern1mu}}
\newcommand{\dx}{{\scriptstyle \Delta} x}
\newcommand{\intdx}{\,\mathrm{d}x}
\newcommand{\mdx}{m\,{\scriptstyle \Delta} x}
\newcommand{\inm}{\left[m\right]}
\newcommand{\inmp}{\left[m^{\prime}\right]}
\newcommand{\eu}[1]{\exp\left[{#1}\right]}

\begin{document}

\title{Basis set convergence of Wilson basis functions for electronic structure}
\author{James Brown}
\author{James D Whitfield}
\affiliation{Department of Physics and Astronomy, Dartmouth College, Hanover, New Hampshire 03755, USA}
\email{james.d.whitfield@dartmouth.edu}

\begin{abstract}
There are many ways to numerically represent of chemical systems in order to compute their electronic structure.  Basis functions may be localized in real-space (atomic orbitals), in momentum-space (plane waves), or in both components of phase-space. Such phase-space localized basis functions in the form of wavelets, have been used for many years in electronic structure.  In this paper, we turn to a phase-space localized basis set first introduced by K. G. Wilson.  We provide the first full study of this basis and its numerical implementation.  To calculate electronic energies of a variety of small molecules and states, we utilize the sum-of-products form, Gaussian quadratures, and introduce methods for selecting sample points from a grid of phase-space localized Wilson basis.  Both full configuration interaction and Hartree-Fock implementations are discussed and implemented numerically. As with many grid based methods, describing both tightly bound and diffuse orbitals is challenging so we have considered augmenting the Wilson basis set as projected Slater-type orbitals.  We have also compared the Wilson basis set against the recently introduced wavelet transformed Gaussians (gausslets).  Throughout, we give comments on the implementation and use small atoms and molecules to illustrate convergence properties of the Wilson basis.
\end{abstract}

\maketitle
\tableofcontents

\section*{Notation}
\begin{enumerate}
    \item $\dx$, Spacing between basis functions in position space.
    \item $m$, Index for location of phase-space basis function in position space at $m\dx$, not necessarily an integer but $m$ has integer spacing.
    \item $k$, Index for location of phase-space basis function in momentum space.
    \item $n$, Signifies phase-space basis function composite index $n={m,k}$.
    \item  $ \vec{n}=\left[n_1,n_2,...n_D\right] $, The vector of all 1D indices for the multidimensional basis function with $D=3$ or $D=6$
    \item $L_k$, Number of grid functions in position space.
    \item $L_m$, Number of grid functions in momentum space.
    \item $L$, Total number of one dimensional phase-space functions.
    \item $N$, Size of basis.
    \item $\bs{H}$, Matrices will be capital letters in bold and non-italicized. 
    \item $\bs{v}$, Vectors will be small letters in bold and non-italicized.
    \item $I(O)$, The integral for the operator $O$.
    \item $G\left(A,B,C,D,\mu,x\right)$, The combination of two Gaussians.
    \item $g\left(A,B,C,D\right)$, The integral of $G\left(A,B,C,D,\mu,x\right)$ over all $x$.
\end{enumerate}

\section{Introduction}

Computation of the electronic structure of a fixed nuclear potential occupies a swath of academic and commercial research in both the quantum and classical computational domains.  In all cases, the basis set used to represent the possible locations of electrons plays a large role in the reach of a finite computational device.  Both plane wave computations and spatially local basis functions form the major approaches.  A key alternative, which includes the present study, makes use of basis functions localized in both momentum and real space.  Such phase-space localized (PSL) basis functions in the form of wavelets have been used in electronic structure calculations with some success for a number of years.\cite{Arias1999,Goedecker1998,Harrison2004,Heinz2002,Cho1993,Flad2006,Genovese2008,Shimshovitz2012b,White2017} However, these basis functions are, with one exception,\cite{White2017} defined only on a grid which makes obtaining analytic matrix elements very difficult for the unbounded Coulomb potential. In 1987, K. G. Wilson introduced a basis which is analytic everywhere and localized in phase-space.\cite{Wilson1987,Sullivan1987}

\Ref{Sullivan1987} was the first to consider Wilson basis functions for electronic structure and suggests that PSL functions should have certain advantages over other grid basis methods.  This early work only considered only a one-dimensional electronic problem and hints at the methodology for performing numerical calculations in real space.  Here we develop the full machinery necessary to utilize Wilson basis sets very similar to the I. Daubechies, S. Jaffard, and J. Journ\cite{Daubechies1991} construction for electronic structure calculations.    The sum-of-products form of the full Hamiltonian in the Wilson basis allows the Hamiltonian to be decomposed into the separate Cartesian directions.  This reduces memory requirements and can be used to perform matrix-vector products more efficiently using sequential summation.\cite{Wodraszka2017} We also use Gaussian quadrature to numerically evaluate the Hamiltonian terms.  Throughout we use a variety of small electronic systems using both exact diagonalization (i.e. full configuration interaction) and the Hartree-Fock approximation.  Further, our paper finds and describes reasonable numerical parameters and procedures for the Wilson basis and its extensions. 

The organization of the paper is as follows. The Wilson basis is introduced in Section \ref{sec.WBF}. The representation of the operators in the Wilson basis is computed in Section \ref{sec.ints} as closed form expressions.  We then describe the evaluation of the Coulomb integrals using Gaussian quadrature in Section \ref{sec.tensor}.  Next, in Section \ref{sec.calcenergies}, we combine the integrals over the Wilson basis with sampling techniques to yield a complete computational procedure for calculating electronic energies. Then the parameters of algorithm are varied and tested on the exactly solvable one-electron hydrogen atomic system in Section \ref{sec.oneelectron}. Section \ref{sec.twoelectron} examines the Wilson basis applied to multiple states of the two-electron helium and molecular hydrogen systems.  Section \ref{sec.HF} discusses how Hartree-Fock calculations with the Wilson basis can be performed by using the methodology developed in the paper.  In the penultimate section, Section \ref{sec.tests}, gausslet basis sets and standard Slater-type orbital basis sets are considered as possibilities to extend the Wilson basis.  We summarize our conclusions and give an outlook in Section \ref{sec.conc}.  

Atomic units ($\hbar=m_e=a_0=4\pi\epsilon_0=1$) are used throughout.

\section{Wilson Basis Functions}\label{sec.WBF}
The Wilson basis functions as defined in \Ref{Daubechies1991} are a product of 1D PSL functions that are of the form
\begin{equation}\label{eq.daubmk}
w_n\left(x\right)=\left\{\begin{array}{ll}
\phi\left(x-\frac{m}{2}\right)&\inm \in 2\,\mathbb{Z},k=0 \\
\sqrt{2}\phi\left(x-\frac{m}{2}\right) \cos\left[2\pi kx\right]& \inm+k\in 2\,\mathbb{Z},k>0 \\
\sqrt{2}\phi\left(x-\frac{m}{2}\right) \sin\left[2\pi k x\right] & \inm+k\in 2\,\mathbb{Z}+1,k>0
\end{array}\right. ,
\end{equation}
where modulated Gaussians are used as functions to generate $\phi\left(x\right)$ such that, 
\begin{equation}
\phi\left(x\right)=\sum_{j,l\in\mathbb{Z}}a_{jl}\exp\left[2\iu l\pi x\right]\left(2v\right)^{1/4}\exp\left[-v \pi (2x-j)^2\right].
\end{equation}
In \Ref{Daubechies1991}, the $a_{jl}$ are found either using a Zak transform or a convergent series in momentum space.

In order to simplify the calculations, we use the technique related to that of \Refs{Poirier2004a,Halverson2012} but most similar to \Ref{Brown2016} to generate a Wilson basis. This involves using a grid of modulated Gaussians defined here as, 
\begin{equation}\label{eq.dmk}
d_n\left(x\right)=\left\{\begin{array}{ll}
\frac{\eta_n}{\sqrt{\dx}}\exp\left[{-\frac{\pi}{2 \dx^2}\left(x-\mdx\right)^2}\right]&\inm \in 2\,\mathbb{Z},k=0 \\
\frac{\eta_n}{\sqrt{\dx}}\exp\left[{-\frac{\pi}{2 \dx^2}\left(x- \mdx\right)^2}\right] \cos\left[\frac{\pi}{\dx}k\left(x-\mdx \right)\right]& \inm+k\in 2\,\mathbb{Z},k>0 \\
\frac{\eta_n}{\sqrt{\dx}}\exp\left[{-\frac{\pi}{2 \dx^2}\left(x-\mdx\right)^2}\right] \sin\left[\frac{\pi}{\dx}k\left(x-\mdx\right)\right] & \inm+k\in 2\,\mathbb{Z}+1,k>0
\end{array}\right. ,
\end{equation}
where 
\begin{equation}
\eta_n=2\left(\frac{1}{2(1+(-1)^{\inm+k}\exp\left[{-\pi k^2}\right])}\right)^{1/2}\left(-\iu\right)^{\inm+k}.
\end{equation}
with $\dx$ being the distance between functions in position space and $\inm$ is either the integer part of $m$ or the integer part of $m-1$, depending on where the $k=0$ are chosen to be placed. For these functions, we have $\Delta x \Delta p=\pi$. Each basis function is localized in position space around $m \dx$ where the $L_m$ possible $m$ values have integer spacing but are in general real. The functions are also localized in momentum space at $\pm k \pi/\dx$ where $k$ is a positive integer such that $k=0,1,2,3,...,L_k-1$. The index $n$ is taken to represent the composite index ${k,\inm}$ and the total 2D phase-space grid that represents one real-space dimension is composed of $L=L_m \times L_k$ functions. Each box in positive momentum space takes up $\pi$ but is combined with the corresponding negative momentum space partner such that the full basis function follows the uncertainty principle and is localized in $2\pi$ of phase-space.

\begin{figure}[ht]
\caption{The symmetric orthogonalized versions of \Eq{eq.dmk} with indices $n=\left(m\in\left[-1,0,1\right], k\in \left[0,1,2\right]\right) $. The checkerboard pattern of symmetry is clear with red functions/boxes representing positive symmetry and black functions/boxes representing negative symmetry. Functions with $m=1$ are lines and triangles, while functions with $m=-1$ are lines and stars. The white boxes at $m=\pm 1,k=0$ indicate that these indices are not in the basis. \label{fig.bfunctions}}
\centering
\includegraphics[width=0.9\textwidth]{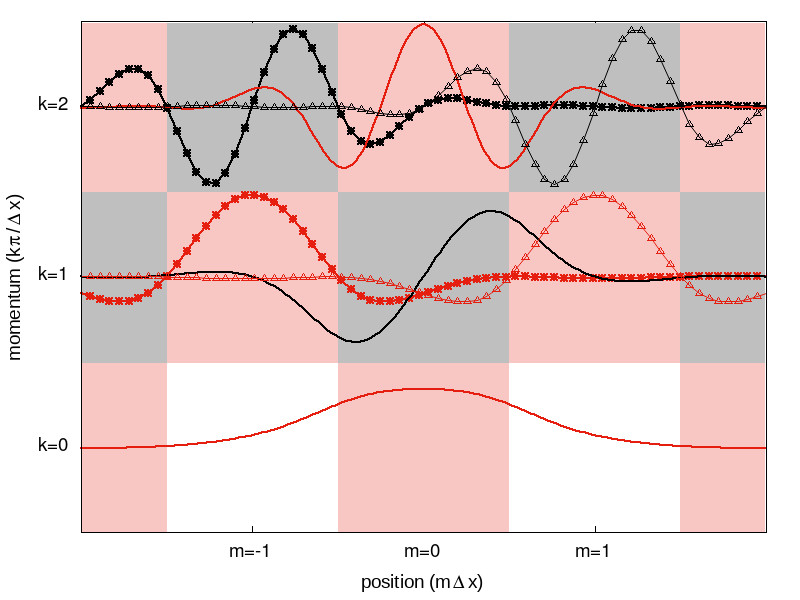}
\end{figure}

\begin{figure}[ht]
\caption{Comparison of the the symmetric orthogonalized versions of \Eq{eq.dmk} ($\tilde{w_n}\left(x\right)$) with $\dx=1/2$, and the DJJ functions $w_n\left(x\right)$ using an exponent scaling factor of $v=1/2$ for positions in phase-space of $n=(m=0,k=1)$ and $n=(m=1,k=1)$. The Gaussian grid from which both sets of functions are generated is a tiling of modulated Gaussians in phase-space spanning indices $m\in \left[-23,23\right],k\in\left[-23,23\right]$. The relative difference between $\tilde{w_n}\left(x\right)$ and $w_n\left(x\right)$ is around $10^{-4}$ and shows a periodicity.  
\label{fig.compfunctions}}
\centering
\includegraphics[width=0.9\textwidth]{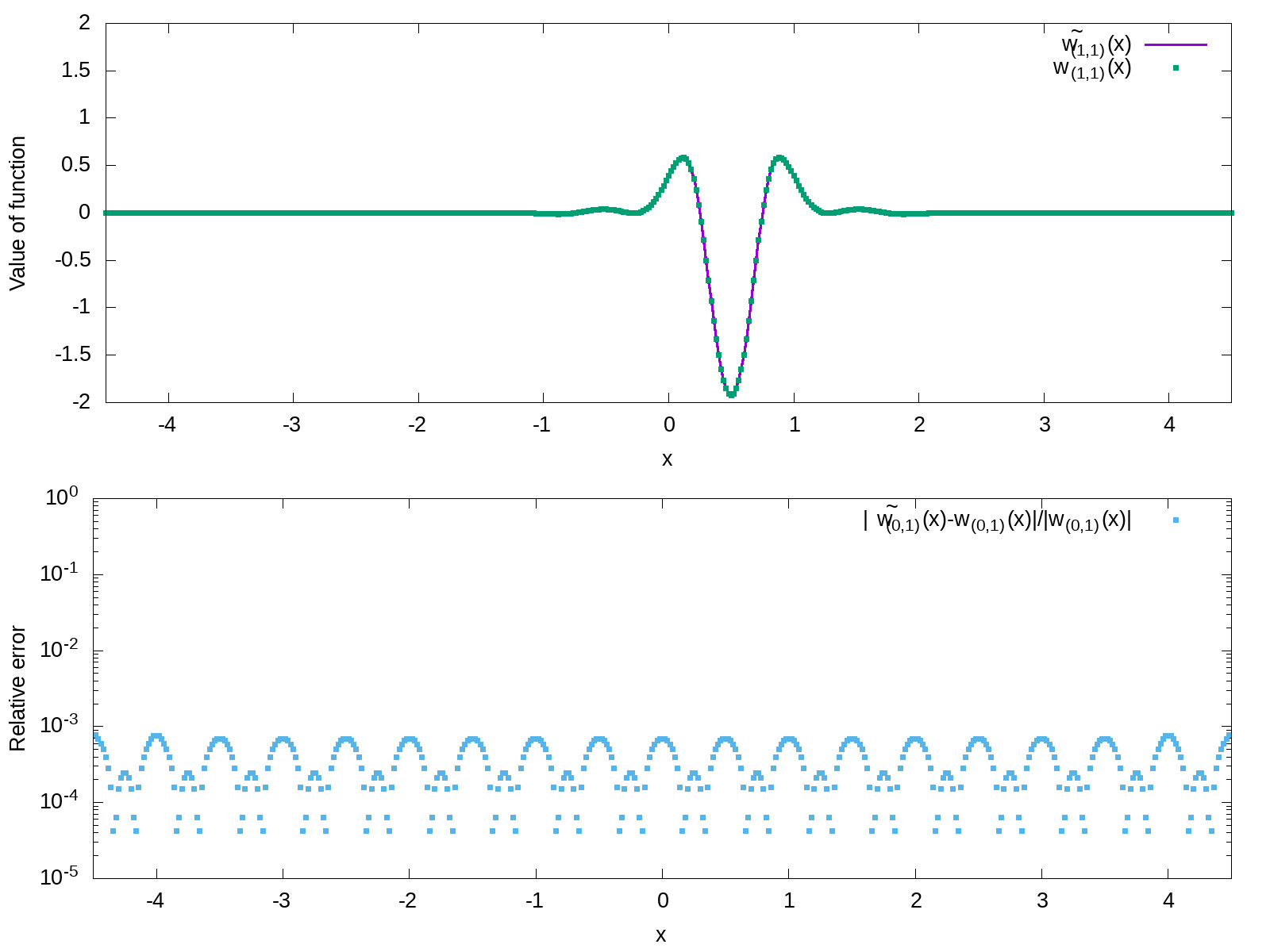}
\end{figure}

As the modulated Gaussians of \Eq{eq.dmk} are not orthogonal to each other, they must be orthogonalized in order to be used for calculating eigenvalues iteratively.  We consider two schemes: symmetric orthogonalized ($\tilde{w}_n\left(x\right)=\sum_j \bs{S}_{nj}^{-1/2}d_j\left(x\right)$) and biorthogonalized ($b_{n}\left(x\right)=\sum_j \bs{S}_{nj}^{-1}d_j\left(x\right) $ combined with $d_{n\p}(x)$ as the dual).  There are benefits to either choice but all resulting basis functions are localized exponentially in phase-space. The symmetric orthogonalized version ($\tilde{w}_n\left(x\right)$)  of these basis functions are shown in  \Fig{fig.bfunctions} and form an orthonormal basis. When the basis functions $\tilde{w}_n\left(x\right)$ are not close to the boundary of the grid, the functions are very similar ($\max_x|\tilde{w}_n(x)-w_n(x)|\lesssim 10^{-3}$) to the I. Daubechies, S. Jaffard, and J. Journ (DJJ) functions $w_n\left(x\right)$.  This can be seen in \Fig{fig.compfunctions} for the $n=\left(0,1\right)$ and $n=\left(1,1\right)$ positions in phase-space. The two depicted functions are representative of all $m,k$ values in the interior of the phase-space domain. Equivalency was claimed in \Ref{Poirier2004a} assuming the underlying modulated Gaussians used to generate the basis functions are the same. However, \Ref{Poirier2004a} did not explicit show the equivalency and also did not appear to rigorously prove it. We do not make an effort to prove the equivalency of $w\left(x\right)$  and $\tilde{w}_n\left(x\right)$ but note that $\tilde{w}_n\left(x\right)$, $b_n\left(x\right)$ and $d_n\left(x\right)$ all have the properties of the original Wilson basis.

All functions ($\tilde{w}_n\left(x\right)$, $b_n\left(x\right)$ and $d_n\left(x\right)$) have symmetry properties in a checker board like fashion where $\inm+k\in 2\,\mathbb{Z}+1$ are odd and $\inm+k\in 2\,\mathbb{Z}$ are even. $\tilde{w}_n\left(x\right)$ and $b_n\left(x\right)$ are exponentially localized in phase-space while $d_n\left(x\right)$ is Gaussian localized. If a smaller grid of $d_n\left(x\right)$ is used to generate $\tilde{w}_n\left(x\right)$, exponential localization is retained but the functions are not as similar (especially for functions at the edge of the grid) to the construction of \Ref{Daubechies1991}. That being said, the accuracy and pruneability of the basis is not impacted greatly.\cite{Halverson2012,Brown2016}

The symmetric (or anti-symmetric) position in momentum space is more explicitly obvious when \Eq{eq.dmk} is in the complex exponential form with $\cos\left[px\right]=1/2\left(\exp\left[{\iu p x}\right]+\exp\left[{-\iu p x}\right]\right)$, $\sin\left[px\right]=-\iu/2\left(\exp\left[{\iu p x}\right]-\exp\left[{-\iu p x}\right]\right)$. This allows the Wilson basis of \Eq{eq.dmk} to be written more succinctly as
\begin{align}\label{eq.cdmk}
    d_n\left(x\right)=&\frac{\eta_n}{\sqrt{\Delta x}}
      \exp\left[{-\frac{\pi}{2 \dx^2}\left(x- \mdx\right)^2}\right]\times
       \nonumber \\ 
      &
    \left(\exp\left[{\iu\frac{\pi}{\dx}k\left(x-\mdx \right)}\right]+(-1)^{[m]+k}\exp\left[{-\iu\frac{\pi}{\dx}k\left(x-\mdx \right)}\right]\right)
\end{align}

In \Eq{eq.cdmk},
the values of $\inm\in 2\mathbb{Z}+1,k=0$ not included in the basis. If one ignored the normalization pre-factor, all $\inm \in 2\mathbb{Z}+1, k=0$ evaluate to zero.

We can further simplify notation by considering each the positive and negative momentum term separately in \Eq{eq.cdmk} using
\begin{equation}
    \alpha_{n}^{\pm}(x)=\frac{\eta_n}{\sqrt{\Delta x}} \exp\left[{-\frac{\pi}{2 \dx^2}\left(x- \mdx\right)^2}\right] \exp\left[{\pm k\frac{\iu\pi}{\dx}\left(x-\mdx \right)}\right]
    \label{eq.cdmka}
\end{equation}
which results in  $d_n(x)=\alpha_n^{+}(x)+(-1)^{\inm+k}\alpha_n^{-}(x)$.

To form a multidimensional basis, a product of the $1D$ basis functions of \Eq{eq.cdmk} for each of the $D$ dimensions is used.  The number of dimensions is either three or six.  In this section we are considering one electron integrals so $D=3$ but when considering the two electron integrals $D=6$. In both cases, the basis function is written as
\begin{equation}\label{eq.Dbasis}
    b_{n_1,n_2,...,n_D}(x_1,x_2,...,x_D)=\prod_{i=1}^D d_{n_i}\left(x_i\right)
\end{equation}
where there is a composite index $n_i$ for each of the $i=1,2,...,D$ dimensions.

\section{Operator representation integrals}\label{sec.ints}
This section outlines all the integrals needed to define the Hamiltonian representation in the Wilson basis. The one and two-body Coulomb integrals can be calculated advantageously as a sum-of-products. 
In electronic structure, the sum-of-product form has been previously used to assist in the evaluation of exchange and Coulomb integrals,\cite{Khoromskij2011} as well as evaluating the M{\o}ller-Plesset perturbation second order correction.\cite{Hohenstein2012}  Exploiting grid based basis sets and a sum-of-product decomposition of the Hamiltonian into its three Cartesian product, has been applied to electronic structure, using a multi-resolution disjoint Legendre polynomial basis,\cite{Harrison2004} and a tensor decomposed sinc function basis.\cite{Jerke2018} Like the Wilson basis, the sinc functions are analytic.  However, the Wilson basis functions are also localized in phase-space which has advantages for representing the cusps of electronic wavefunctions.

Due to the sum-of-products form, the overlap matrix and the Cartesian kinetic energy operator can be defined using a product of $1D$ matrix elements.  All operator $O$ integrals will be performed using the notation of $I^{(a\p a)}_{n\p n}\left(O\right)$ which refers to the pair $\alpha_{n\p}^{a\p},\alpha_{n}^{a}$ with $a\p, a\in \{+,-\}$ except for the overlap integral denoted $S^{(a\p a)}_{n\p n}$.  The integral then includes four pairs of momentum combinations which are $(+k\p,+k)$, $(+k\p,-k)$, $(-k\p,+k)$, and $(-k\p,-k)$.  The $I^{(+,+)}_{n\p n}\left(O\right)$ and $I^{(+,-)}_{n\p n}\left(O\right)$ integrals are equivalent to the complex conjugates of  the $I^{(-,-)}_{n\p n}\left(O\right)$ and $I^{(-,+)}_{n\p n}\left(O\right)$ integrals respectively.  The two-electron terms have a similar structure but require summing over more momentum combinations.

The overlap matrix for two basis functions in three dimensions is $\vec{n}\p=\left[n_x\p,n_y\p,n_z\p\right]$ and $\vec{n}=\left[n_x,n_y,n_z\right]$ is 
\begin{equation}
S_{\vec{n}\p\vec{n}}=\prod_{i}S_{n_i\p n_i}= 
\int \mathrm{d}x \, d_{n_x}(x)d_{n\p_x}(x) 
\int \mathrm{d}y \, d_{n_y}(y)d_{n\p_y}(y) 
\int \mathrm{d}z \, d_{n_z}(z)d_{n\p_z}(z) 
\end{equation}
while the portion of the Cartesian Laplacian with second derivative in dimension $j$ is written as
\begin{equation}
T^{(j)}_{\vec{n}\p\vec{n}}=T_{n_j\p,n_j}\prod_{i\neq j}S_{n_i\p n_i}.
\end{equation}

Throughout we will be using Gaussian integrals heavily so let us establish notation. We will always represent Gaussian integrals in the form
\begin{equation}\label{eq.start1}
g\left(A,B,C,D\right)=\int_{-\infty}^{\infty}\mathrm{d}x\, G\left(A,B,C,D,\mu,x\right)
\end{equation} 
where
\begin{equation}\label{eq.start2}
G\left(A,B,C,D,\mu,x\right)= \exp\left[-A \left(x-\mu\right)^2\right]\exp[\iu B \left(x-\mu \right)]\exp[C]\exp[\iu D]
\end{equation} 
with $A,B,C,D,\mu$ are real constants. The integration of \Eq{eq.start1} results in
\begin{equation}\label{eq.sol}
g\left(A,B,C,D\right)= \left(\frac{\pi}{A}\right)^{1/2}\exp[-\frac{B^2}{4 A}]\exp[C]\exp[\iu D]
\end{equation}

The integrals of $x$ and $x^2$ are also presented as they are used to efficiently calculate the $\nabla^2$ operator below.
\begin{equation}
    \int_{-\infty}^{\infty}\mathrm{d}x\, x\, G\left(A,B,C,D,\mu,x\right)=\left(i\frac{B}{2A}+\mu\right)g(A,B,C,D)
    \label{eq.xG}
\end{equation}
and 

\begin{equation}
    \int_{-\infty}^{\infty}\mathrm{d}x\, x^2\, G\left(A,B,C,D,\mu,x\right)=\left(\frac{2A-B^2}{4A^2}+\iu \mu \frac{B}{A}+\mu^2\right) g(A,B,C,D)
    \label{eq.x2G}
\end{equation}

\subsection{Overlap Integral}\label{sec.overlap}
The overlap integral that needs to be evaluated for all four $(a \p,a)$ combinations is given as
\begin{eqnarray}\label{eq.int1}
\tilde{S}^{(a\p a)}_{n\p n}&=&\int_{-\infty}^\infty dx\, \alpha_{n'}^{a'}(x)\alpha_{n}^{a}(x)=\left(\eta_n \eta_{n^{\prime}}\frac{1}{\dx}\right)\int_{-\infty}^\infty \intdx\, G(A,B,C,D,\mu,x)\nonumber \\
&=&
\int_{-\infty}^\infty \left(\eta_{n^{\prime}}\eta_n \frac{1}{\dx}\right)\exp\left[{-\frac{\pi}{2 \dx^2}\left(x-m\p \dx\right)^2}\right]\exp\left[{\iu \frac{\pi}{\dx}a\p k\p \left(x-m\p \dx\right)}\right]\times \nonumber\\
&&\exp\left[{-\frac{\pi}{2 \dx^2}\left(x-\mdx\right)^2}\right]\exp\left[{\iu \frac{\pi}{\dx}a k\left(x-\mdx\right)}\right]\intdx,
\end{eqnarray}
where either $(-1)^{\inm+k}$ or $(-1)^{[m\p]+k\p}$ are excluded from $\tilde{S}^{(a\p a)}_{n\p n}$ but are included in the definition of $S_{n\p n}^{(a\p a)}$. 

For \Eq{eq.int1}, the appropriate values of $A,B,C,D$ for the product of $\alpha_{n}^{a}(x)$ and $\alpha_{n'}^{a'}(x)$ are
\begin{align}\label{eq.wilsonABCD} 
A&=\frac{\pi}{\dx^2},&
B&=\frac{\pi}{\dx}k^{(a\p a)}_- ,&
C&=-\frac{\pi}{4} m_-^2,&
D&=\frac{1}{2}\left(m_- k^{(a\p a)}_+\right)\pi,&
\mu&=\frac{\dx}{2}m_+.
\end{align}
where $m_\pm=m\p\pm m$ and $k^{(a\p a)}_\pm=a\p k\p\pm ak$.
Thus, the overlap integral for a given $a\p, a$ combination evaluates to
\begin{equation}
\begin{split}
\tilde{S}_{n\p  n}^{(a\p a)}&=\left(\eta_n \eta_{n^{\prime}}\frac{1}{\dx}\right)g(A,B,C,D,\mu)\\
&=\eta_n \eta_{n^{\prime}}\exp\left[-\frac{\pi}{4}\left(m_-^2+\left(k^{(a\p a)}_+\right)^2\right)\right] \exp\left[-\iu \frac{\pi}{2}\left(m_- k^{(a\p a)}_-\right) \right].
\end{split}
\end{equation}
Each of the $a\p,a$ combinations are added together to form the full overlap matrix element (i.e. $S_{n\p  n}=\tilde{S}_{n\p  n}^{(++)}+(-1)^{\inm+k+\inmp+k\p}\tilde{S}_{n\p  n}^{(--)}+(-1)^{\inm+k}\tilde{S}_{n\p  n}^{(+-)}+(-1)^{\inmp+k\p}\tilde{S}_{n\p  n}^{(-+)}$).  The final definition of the partial overlap values is
\begin{equation}
\begin{array}{ccl}
S_{n\p  n}^{(++)}&=&\tilde{S}_{n\p  n}^{(++)}\\
S_{n\p  n}^{(+-)}&=&(-1)^{\inm+k}\tilde{S}_{n\p  n}^{(+-)}\\
S_{n\p  n}^{(-+)}&=&(-1)^{\inmp+k\p}\tilde{S}_{n\p  n}^{(-+)}\\
S_{n\p  n}^{(--)}&=&(-1)^{\inm+k+\inmp+k\p}\tilde{S}_{n\p  n}^{(--)}\label{eq.S4}
\end{array}
\end{equation}
where the appropriate $(-1)^{\inm+k}$ or $(-1)^{\inm\p+k\p}$ are now included. The full overlap matrix for basis functions $d_n(x)$ with $d_{n\p}(x)$ is 
\begin{equation}\label{eq.fullover}
S_{n\p  n}=\sum_{a}^{+,-}\sum_{a\p}^{+,-}{S}_{n\p  n}^{(a\p a)}.    
\end{equation}

\subsection{1D Kinetic energy operator}
All 1D operators examined here can be written as
\begin{equation}\label{eq.intform}
I_{n\p  n}(O)=\sum_{a}^{+,-}\sum_{a\p}^{+,-}I^{(a\p a)}_{n\p  n}(O)
\end{equation}
where $I^{(a\p a)}_{n\p  n}(O)$ are the partial integrals for operator $O$. 

When evaluating the integrals associated with the kinetic energy operator, the values of \eqref{eq.wilsonABCD} will be the same since the kinetic energy operator applied to a complex Gaussian results in the same Gaussian multiplied by a second degree polynomial.  Therefore we will also need to integrate against the $x$ and $x^2$ operators to evaluate the Laplacian.

Using \eqref{eq.xG} and \eqref{eq.x2G}, the necessary partial integrals are given in terms of the overlap integral as
\begin{eqnarray}
I^{(a\p a)}_{n\p  n}(x)&=&\left(i \frac{B}{2A}+ \mu\right)S^{(a\p a)}_{n\p  n} \nonumber \\[1ex]
&=& \frac{\dx}{2}\left(m_+ +\iu (a\p k\p+a k)\right)S^{(a\p a)}_{n\p  n}
\end{eqnarray}
and
\begin{eqnarray}
I^{(a\p a)}_{n\p  n}(x^2)&=&\left(\frac{2A-B^2}{4A^2}+\iu \mu \frac{B}{A}+\mu^2\right) S^{(a\p a)}_{n\p  n}\nonumber \\[1ex]
&=&\frac{\dx^2}{4}\left(\frac{2}{\pi}+m_+^2 + \iu 2  m_+ (a\p k\p + a k)- (a\p k\p+ a k)^2\right)S^{(a\p a)}_{n\p  n}
\end{eqnarray}
where $S^{(a\p a)}_{n\p  n}$ are defined in \Eq{eq.S4} with $a,a\p \in \left\{+,-\right\}$. 

To evaluate the integral of the $-\nabla^2$ operator, we will use the Cartesian representation $(-\frac{\mathrm{d}^2}{\mathrm{d}x^2}-\frac{\mathrm{d}^2}{\mathrm{d}y^2}-\frac{\mathrm{d}^2}{\mathrm{d}z^2})$ and perform 1D integrals for $x,y,z$. For greater symmetry in the integrals, we will act with the derivative operator to the left and to the right with form  $\frac{\mathrm{d}}{\mathrm{d}x}^{\dagger}\frac{\mathrm{d}}{\mathrm{d}x}\equiv-\frac{\mathrm{d}^2}{\mathrm{d}x^2}$. The action of $\frac{\mathrm{d}}{\mathrm{d}x}^{\dagger}\frac{\mathrm{d}}{\mathrm{d}x}$ in the integral simplifies to,
\begin{eqnarray}
\left(\frac{\mathrm{d}}{\mathrm{d}x}\alpha_{n\p}^{a\p} \right)\times\left( \frac{\mathrm{d}}{\mathrm{d}x}\alpha_{n}^{a}\right)\nonumber
&=&\Bigg(\frac{\pi^2}{4 \dx^2}\left(2 m\p - \iu 2 a\p k\p\right)\left(2m + \iu 2 a k\right)-\\&&
\frac{\pi}{\dx^3}\left(-\iu m_+ +k^{(a\p a)}_+\right)x+
\frac{\pi^2}{\dx^4} x^2\Bigg)
 \times G(A,B,C,D,\mu,x)
 \end{eqnarray}
Using the $S^{(a\p a)}_{n\p n},I^{(a\p a)}_{n\p n}(x)$ and $I^{(a\p a)}_{n\p n}(x^2)$ derived above and simplifying results in
\begin{equation}\label{eq.keoint}
I^{(a\p a)}_{n\p n}\left(\frac{\mathrm{d}}{\mathrm{d}x}^{\dagger}\frac{\mathrm{d}}{\mathrm{d}x}\right)=\frac{\pi^2}{4 \dx^2}\left(\frac{2}{\pi}-\left(m_-+\iu k^{(a\p a)}_-\right)^2 \right)S^{(a\p a)}_{n\p n}.
\end{equation}
One can then use \Eq{eq.intform} to obtain the complete integral by summing over the four possible values of $a\p, a$, to obtain
\begin{equation}\label{eq.fullkeo}
T_{n\p n}=\sum_{a\p,a}^{+,-}I^{(a\p a)}_{n\p n}\left(\frac{\mathrm{d}}{\mathrm{d}x}^{\dagger}\frac{\mathrm{d}}{\mathrm{d}x}\right)    
\end{equation}
which can be stored in an $L\times L$ matrix $\bs{T}$ and is the same for $x,y,z$.

\subsection{Coulomb integral}
The Coulomb integral is more challenging and will be the focus of the remainder of this section.  We evaluate the integrals in closed form using the error function, $\textrm{erf}(x)=2\pi^{-1/2}\int_0^x\exp(-t^2)dt$, as well as discuss the use of numerical quadrature to evaluate the functions quickly.  The numerical integration is, in the end, preferred since it allows the integration to be done component-wise.  As we will see below, the input to the error function has all Cartesian components combined.

To exploit the fact that our multi-dimensional basis functions, \Eq{eq.Dbasis}, are products of each Cartesian dimension, the form of the Coulomb operator we use is
\begin{equation}\label{eq.coun}
\frac{1}{|r-r'|}=\frac{2}{\sqrt{\pi}}\int_0^{\infty} \exp\left[{-t^2(x-x')^2}\right] \exp\left[{-t^2(y-y')^2}\right] \exp\left[{-t^2(z-z')^2}\right]\mathrm{d} t
\end{equation}
where $r=[x,y,z]$ and $r'=[x',y', z']$. We will use \Eq{eq.coun}, to perform the electron-nuclear and electron-electron integrations in close form.

\subsubsection{Electron-nuclear Coulomb integral}
The electron-nuclear integrals are performed using $r'=r_A=[R_x,R_y,R_z]$ as the fixed nuclear position.  We will evaluate the integral using the form \Eq{eq.coun}; first by integrating over the one-dimensional coordinates, then by integrating over the dummy variable $t$ of \Eq{eq.coun}.  Since all terms including those introduced by the integral form of the Coulomb operator are all Gaussian, there are many simplifications along the way.  

The required integral is given by
\begin{equation}
I_{\vec{n}'\vec{n}}^{\vec{a}\p,\vec{a}} \left(\frac{1}{|r-r\p|}\right)= 
\int_{-\infty}^{\infty} \frac{G_x(x)G_y(y)G_z(z)}{|r-r'|} dx dy dz
\end{equation}
The constants for $G_j(r)=G(A_j,B_j,C_j,D_j,\mu_j,r)$ are the same as those given in \Eq{eq.wilsonABCD}.  In general, these constant will depend on which direction is being discussed, hence the subscript.

We can simplify notation here and in subsequent sections by performing the integral over the individual spatial coordinates before simplifying the Gaussian expression which then remains parameterized by $t$.  Consider for $j=x,y,z$ the integration
\begin{eqnarray}
\widetilde{V}_{n_j\p n_j}^{(a_j\p a_j)}(t)&=&\int_{-\infty}^{\infty}\mathrm{d}q\,G_j(q) \exp[-t^2 (q-R_j)^2]=g(A_{jt},B_{jt},C_{jt},D_{jt})\label{eq.Vt}
\end{eqnarray}
The constants for the final Gaussian integral are given by
\begin{align}
A_{jt}&=A_j+t^2,&
B_{jt}&=B_j&\nonumber\\
C_{jt}&=C_j-\frac{\pi}{4}\frac{\left(\dx \, m_{j+} -2 R_i\right)t^2}{\pi+\dx^2 t^2}&
D_{jt}&=D_j+\frac{\pi}{2}\frac{\left(2R_i- \dx \,m_+\right)t^2}{\pi+\dx^2 t^2}
\label{eq.counabcd}
\end{align}
Anticipating the final integration, we perform a change of variables to further convert the integral over $t\in(0,\infty)$ to $v\in(-1,1)$ with \begin{equation}
    t=\sqrt{\frac{\pi}{\dx^2}\frac{1-v}{1+v}},\quad
    \mathrm{d}t=-\frac{\sqrt{\pi}}{\dx}\frac{1}{\sqrt{1-v^2}}\mathrm{d}v
\end{equation}
Performing the change of variables in \Eq{eq.Vt} and integrating over $q$ results in
\begin{eqnarray}
\widetilde{V}^{(a_i\p a_i)}_{n_i\p n_i}(v,R_i)&=& \sqrt{\frac{\left(1+v\right)}{2}} \times \eu{- (c_{n_i\p n_i}^{(a_i\p a_i)})\left(1-v\right)}S_{n_i\p n_i}^{(a_i\p a_i)} 
\end{eqnarray} 
where $c_{n_i\p n_i}^{(a_i\p a_i)}=\frac{\pi}{8}(m_{i+} + \iu k_{+}^{(a\p a)}-2\frac{R_i}{\dx})^2$. Our final definition for $V$ removes the prefactor
\begin{equation}
V^{(a_i\p a_i)}_{n_i\p n_i}(v,R_i)=\sqrt{\frac{2}{\left(1+v\right)}}\widetilde{V}^{(a_i\p a_i)}_{n_i\p n_i}(v,R_i)=\eu{- (c_{n_i\p n_i}^{(a_i\p a_i)})\left(1-v\right)}S_{n_i\p n_i}^{(a_i\p a_i)} 
\end{equation}

The full 3D integral for a given $\vec{n},\vec{a},\vec{n}\p,\vec{a}\p$ is then reduced to
\begin{equation}\label{eq.sopcoul}
I_{\vec{n}'\vec{n}}^{\vec{a}\p,\vec{a}} \left(\frac{1}{|r-r\p|}\right)=\frac{1}{\dx}\frac{1}{\sqrt{2}}\int_{-1}^{1}\mathrm{d}v\,\frac{ V^{(a_x\p a_x)}_{n_x\p n_x}(v,R_i)V^{(a_y\p a_y)}_{n_y\p n_y}(v,R_i)V^{(a_z\p a_z)}_{n_z\p n_z}(v,R_i)}{\sqrt{1-v}}.
\end{equation}

The complete integral for a nucleus at $r\p$ for a basis function $\vec{n}\p=\left[n_x\p,n_y\p,n_z\p\right]$ with $\vec{n}=\left[n_x,n_y,n_z\right]$ is then 
\begin{equation}\label{eq.coul_nuc}
I_{\vec{n}\p\vec{n}}\left(\frac{1}{r-r_e}\right)=\sum_{\{a\}}^{64} \frac{1}{\dx}\frac{1}{\sqrt{2}}\int_{-1}^{1}\mathrm{d}v\,\frac{ V^{(a_x\p a_x)}_{n_x\p n_x}(v,R_x)V^{(a_y\p a_y)}_{n_y\p n_y}(v,R_y)V^{(a_z\p a_z)}_{n_z\p n_z}(v,R_z)}{\sqrt{1-v}}
\end{equation}
where the sum is performed over all $64$ combinations of $a_x, a_y, a_z, a_x',a_y', a_z'$ which can each be plus or minus. Evidently, the 3D Coulomb integral can be written in the multidimensional form of \Eq{eq.intform} with an additional integration over variable $v$. \Eq{eq.coul_nuc} can be evaluated in closed form to,
\begin{equation}\label{eq.coul_nuc_erf}
I_{\vec{n}\p\vec{n}}\left(\frac{1}{r-r_e}\right)=\sum_{\{a\}}^{64} \frac{\sqrt{\pi}}{\dx} \frac{\mbox{Erf}\left[\sqrt{ c_{n_x\p n_x}^{(a_x\p a_x)}+c_{n_y\p n_y}^{(a_y\p a_y)}+c_{n_z\p n_z}^{(a_z\p a_z)}}\right]}{\sqrt{ c_{n_x\p n_x}^{(a_x\p a_x)}+c_{n_y\p n_y}^{(a_y\p a_y)}+c_{n_z\p n_z}^{(a_z\p a_z)}}}S_{n_x\p n_x}^{(a_x\p a_x)} S_{n_y\p n_y}^{(a_y\p a_y)} S_{n_z\p n_z}^{(a_y\p a_y)}. 
\end{equation}

\subsubsection{Two-electron Coulomb integral}
The form of the Coulomb operator for the two-electron integral is similar to \Eq{eq.sopcoul} but defining
$\vec{n_i}=\left[n_{x_i},n_{y_i},n_{z_i}\right]$ and $\vec{a_1}=\left[a_{x_i},a_{y_i},a_{z_i}\right]$ is,
\begin{equation}
I_{\vec{n_1}'\vec{n_2}'\vec{n_1}\vec{n_2}}^{a_1\p,a_2\p,a_1,a_2} \left(\frac{1}{|r-r\p|}\right)=\frac{2}{\sqrt{\pi}}\int_{0}^{\infty}\mathrm{d}t\, F_x(t)F_y(t)F_z(t)
\end{equation}
where 
\begin{equation}
F_i(t)=\int_{-\infty}^{\infty}\mathrm{d}q_2\int_{-\infty}^{\infty}\mathrm{d}q_1\,
G_i^{1}(q_1) G_i^{2}(q_2)\eu{-t^2(q_1-q_2)^2}
\end{equation}
Here 
$G_i^{k}=G(A_{i},B_{i},C_{i},D_{i},\mu_i,q_k)$ is the product of Gaussian corresponding to different dimensions $i=x,y,z$ and a one-dimensional electron coordinate $k=1,2$ along that that direction.
To use this function we need to perform the integrals over $x_1,y_1,z_1,x_2,y_2,z_2$ and then perform the integral over $t$. The change of variables here is $v=\frac{\pi-2\dx^2t^2}{\pi+2 \dx^2 t^2}$ which results in
\begin{equation}
\begin{split}
I_{\vec{n_1}'\vec{n_2}'\vec{n_1}\vec{n_2}}^{(a_1\p a_2\p a_1 a_2)}(\frac{1}{r_1-r_2})=\frac{1}{2\dx}\int_{-1}^1 &\mathrm{d}v\,\frac{V_{\vec{n_{x_1}}' n_{x_2}' n_{x_1}n_{x_2}}^{(a_{x_1}\p a_{x_2}\p a_{x_1} a_{x_2})}(v)V_{\vec{n_{y_1}}' n_{y_2}' n_{y_1}n_{y_2}}^{(a_{y_1}\p a_{y_2}\p a_{y_1} a_{y_2})}(v)V_{\vec{n_{z_1}}' n_{z_2}' n_{z_1}n_{z_2}}^{(a_{z_1}\p a_{z_2}\p a_{z_1} a_{x_2})}(v)}{\sqrt{1-v}} 
\end{split}
\end{equation}
where
\begin{equation}
    V_{\vec{n_{i_1}}' n_{i_2}' n_{i_1}n_{i_2}}^{(a_{i_1}\p a_{i_2}\p a_{i_1} a_{i_2})}(v)=\eu{-b_i^{(a_1\p,a_1,a_2\p,a_2)}\left(1-v\right)}S^{(a_{i_1}\p a_{i_1})}_{n_{i_1}\p,n_{i_1}}S^{(a_{i_2}\p a_{i_2})}_{n_{i_2}\p,n_{i_2}}
\end{equation}
where $b_i^{(a_1\p,a_1,a_2\p,a_2)}=\frac{\pi}{16}\left(m_{i_1+}-m_{i_2+}+\iu(k_{i_1+}^{(a_{i_1}\p a_{i_1})}-k_{i_2+}^{(a_{i_2}\p a_{i_2})})\right)^2$ with $i=x,y,z$.
 
The full 6-dimensional integral is given as a sum of $4^6=4096$ terms in closed form as
\begin{equation}
\begin{split}
I_{\vec{n_1}'\vec{n_2}'\vec{n_1}\vec{n_2}}(\frac{1}{r_1-r_2})=&\sum_{\vec{a}}^{4096}\frac{\sqrt{\pi}}{\dx\, 2\sqrt{2}}\frac{\mbox{Erf}\left[\sqrt{2\left(b_x^{(a_{x_1}\p,a_{x_1},a_{x_2}\p,a_{x_2})}+b_y^{(a_{y_1}\p,a_{y_1},a_{y_2}\p,a_{y_2})}+b_z^{(a_{z_1}\p,a_{z_1},a_{z_2}\p,a_{z_2})}\right)}
\right]}{\sqrt{b_x^{(a_{x_1}\p,a_{x_1},a_{x_2}\p,a_{x_2})}+b_y^{(a_{y_1}\p,a_{y_1},a_{y_2}\p,a_{y_2})}+b_z^{(a_{z_1}\p,a_{z_1},a_{z_2}\p,a_{z_2})}}}\times \\ & S^{(a_{x_1}\p a_{x_1})}_{n_{x_1}\p,n_{x_1}}S^{(a_{y_1}\p a_{y_1})}_{n_{y_1}\p,n_{y_1}}S^{(a_{z_1}\p a_{z_1})}_{n_{z_1}\p,n_{z_1}}S^{(a_{x_2}\p a_{x_2})}_{n_{x_2}\p,n_{x_2}}S^{(a_{y_2}\p a_{y_2})}_{n_{y_2}\p,n_{y_2}}S^{(a_{z_2}\p a_{z_2})}_{n_{z_2}\p,n_{z_2}}.
\end{split}
\end{equation}
Note that $\Delta x$ only appears as a prefactor to the integral.  Thus, the Coulomb integrals do not need to be re-evaluated when changing the spatial distance between adjacent Wilson basis functions in real space.

\section{Evaluating Coulomb matrix elements using Gaussian quadrature}\label{sec.tensor}

In the previous section, we have evaluated all the overlap, kinetic and Coulomb integrals in closed form using the error function.  However, the error function is not separable since the $x,y,z$ components are combined in the argument of the error function. In order to use a much larger number of basis functions, it is important to a) not store the full matrix b) increase the efficiency of matrix-vector products.  

We can accomplish this using Gaussian quadrature over each dimension separately to get the Hamiltonian into a sum-of-products form.  The overlap integral and the kinetic energy integrals are already a sum-of-products of one-dimensional operators so we only need to consider the Coulomb integrals.

Let us consider integrals of the form \Eq{eq.coul_nuc} 
which is in the form $\int_{-1}^1 f(x) (1-x)^a (1+x)^b$ with $a=-0.5$ and $b=0$.  Integrals of this form can be evaluated using Gauss-Jacobi quadrature \cite{Trefethen2013} where the integral is approximated using $\sum^{N_p}_\alpha w_\alpha f(p_\alpha)$. Here $N_p$ is the number of quadrature points $p_{\alpha}$ and their corresponding weights are $w_{\alpha}$. 

This allows us to accurately convert the integral \Eq{eq.sopcoul} into a sum-of-products form with $N_p$ terms such that,
\begin{equation}\label{eq.sop}
I_{\vec{n}\p\vec{n}}\left(\frac{1}{r-r_c}\right)\approx \sum_{\alpha=1}^{N_p}w_{\alpha} V_{n_x\p n_x}^{(1)}(p_{\alpha},R_x) V_{n_y\p n_y}^{(1)}(p_{\alpha},R_y) V_{n_z\p n_z}^{(1)}(p_{\alpha},R_z)
\end{equation} 
where
\begin{equation}\label{eq.fulloneelec}
V_{n_i\p n_i}^{(1)}(p_{\alpha},R_i)=\sum_{a_i\p,a_i}^{+,-}V^{(a_i\p a_i)}_{n_i\p n_i}(p_{\alpha},R_i),
\end{equation}
with $i=x,y,z$. $\bs{V^{(1)}}(p_{\alpha},R_i)$ is an $L \times L$ matrix with each element being a sum of $2^2=4$ terms.

Since Gaussian quadrature calculates polynomials of order $2N_p+1$ exactly, the convergence of the integral is superb. Test calculations have found that as few as seven quadrature points are necessary to obtain chemical accuracy as shown in the test results below. The same transformation can be performed for the two-body Coulomb as
\begin{equation}
I_{\vec{n}\p\vec{n}}(\frac{1}{r_1-r_2})\approx \sum_{\alpha=1}^{N_p}\frac{w_{\alpha}}{2}\, 
 V^{(2)}_{n_{x_1}\p n_{x_1}n_{x_2}\p n_{x_2}}(p_{\alpha})
 V^{(2)}_{n_{y_1}\p n_{y_1}n_{y_2}\p n_{y_2}}(p_{\alpha})
 V^{(2)}_{n_{z_1}\p n_{z_1}n_{z_2}\p n_{z_2}}(p_{\alpha})
\end{equation} 
where 
\begin{equation}\label{eq.fulltwoelec}
V^{(2)}_{n_{i_1}\p n_{i_1}n_{i_2}\p n_{i_2}}(p_{\alpha})=\sum_{a_{i_1},a_{i_2},a_{i_1}\p,a_{i_2}\p}^{+,-}V_{\vec{n_{i_1}}' n_{i_2}' n_{i_1}n_{i_2}}^{(a_{i_1}\p a_{i_2}\p a_{i_1} a_{i_2})}(p_{\alpha}),
\end{equation}
with $i=x,y,z$. $\bs{V^{(2)}}(p_{\alpha})$ an $L^2 \times L^2$ matrix with each element being a sum of $4^2=16$ terms. It is not possible to separate $x_1,x_2$ in any obvious fashion analytically. This means that storing the two-electron matrix elements grows rapidly with the size of 1D sampling grid, and is responsible for most of the memory usage. The $\bs{V^{(2)}}(p_{\alpha})$ are the same for $x,y,z$ so only one matrix for each of the $N_p$ quadrature points need to be stored in memory.

The main value of these transformations is that we can now make use of the well established technique of performing sums sequentially, which is of common usage in the calculation of vibrational energy levels. The advantage is only realized for a ``normal'' (i.e. not generalized) eigenvalue problem. Using the modulated Gaussian grid of \Eq{eq.cdmk} directly results in a generalized eigenproblem unless we orthogonalize or biorthogonalize the representation. For the one-electron case, the Hamiltonian can be written in a sum-of-products form as
\begin{equation}
H_{\vec{n}\p n}= \sum_{t}^{N_t}\prod_{i=x,y,z}O^{(t)}_{n_i\p n_i}
\end{equation}
where $O^{(t)}_{n_i\p n_i}$ is a 1D operator which is either the kinetic energy integral (of \Eq{eq.fullkeo}), the overlap integral (of \Eq{eq.fullover}) or one of the
Coulomb terms (of \Eq{eq.fulloneelec}). The number of terms in the Hamiltonian representation is $N_t=3+N_n N_p$ with factor of three for the kinetic energy operator directions, $N_n$ is the number of nuclei, 
and $N_p$ is the number of quadrature points used to evaluate each nuclei-electron Coulomb term. Each $\bs{O_{i}^{(t)}}$ for dimension $i$ is of size $L\times L$.
This means that the full generalized eigenvalue problem can be written as
\begin{equation}\label{eq.geneig}
\left(\sum_{t}^{N_t} \bs{O}^{(t)}_x\otimes \bs{O}^{(t)}_{y} \otimes \bs{O}^{(t)}_{z}\right)\bs{Z}=\left(\bs{S}_{x}\otimes \bs{S}_{y} \otimes \bs{S}_{z}\right)\bs{Z}\bs{E}
\end{equation}
where $\bs{Z}$ is the matrix of eigenvectors, and $\bs{E}$ is the diagonal matrix of eigenvalues. The total memory requirements to store the sum-of-products matrix representation is $3N_p L^2$ which is much less than the full matrix which requires $L^6$. The $\bs{Z}$ and $\bs{E}$ are of size $L^3\times L^3$. \Eq{eq.geneig} can easily be converted to an eigenvalue problem with no overlap matrix by either using 
\begin{equation}\label{eq.eigform1}
\left(\sum_{t} \bs{S}^{-1/2}\bs{O}^{(t)}_x\bs{S}^{-1/2}\otimes \bs{S}^{-1/2}\bs{O}^{(t)}_{y}\bs{S}^{-1/2} \otimes \bs{S}^{-1/2}\bs{O}^{(t)}_{z}\bs{S}^{-1/2}\right)\bs{U}=\bs{U}\bs{E}
\end{equation}
with $\bs{U}=(\bs{S}^{1/2}\otimes\bs{S}^{1/2}\otimes\bs{S}^{1/2})\bs{Z}$ or 
\begin{equation}\label{eq.eigform2}
\left(\sum_{t} \bs{O}^{(t)}_x\bs{S}^{-1}\otimes \bs{O}^{(t)}_{y}\bs{S}^{-1} \otimes \bs{O}^{(t)}_{z}\bs{S}^{-1}\right)\bs{U}=\bs{U}\bs{E}
\end{equation}
with $\bs{U}=(\bs{S}\otimes\bs{S}\otimes\bs{S})\bs{Z}$. Both forms have been used successfully in previous vibrational calculations.\cite{Brown2015b} \Eq{eq.eigform1} has the advantage of being Hermitian while \Eq{eq.eigform2} generally produces a smaller basis representation. Most calculations in this study use \Eq{eq.eigform1} but both are examined. We also emphasize that \Eq{eq.eigform1} will result in a calculation very similar to using the Wilson basis of \Ref{Daubechies1991} if a large grid of \Eq{eq.cdmk} is used to generate the $\bs{O_i}^{(t)}$ and $\bs{S}$ matrices. 

Similarly, the two-body Coulomb term can be orthogonalized as 
\begin{equation}\label{eq.eigformc1}
\left(\sum_{t} \bs{S}_{1,2}^{-1/2}\bs{O}^{(t)}_{x_1,x_2}\bs{S}_{1,2}^{-1/2}\otimes \bs{S}_{1,2}^{-1/2}\bs{O}^{(t)}_{y_1,y_2}\bs{S}_{1,2}^{-1/2} \otimes \bs{S}_{1,2}^{-1/2}\bs{O}^{(t)}_{z_1,z_2}\bs{S}_{1,2}^{-1/2}\right)\bs{U}=\bs{U}\bs{E}
\end{equation}
or biorthogonalized as
\begin{equation}\label{eq.eigformc2}
\left(\sum_{t} \bs{O}^{(t)}_{x_1,x_2}\bs{S}_{1,2}^{-1}\otimes \bs{O}^{(t)}_{y_1,y_2}\bs{S}_{1,2}^{-1} \otimes \bs{O}^{(t)}_{z_1,z_2}\bs{S}_{1,2}^{-1}\right)\bs{U}=\bs{U}\bs{E}
\end{equation}
where $\bs{O}^{(t)}_{x_1,x_2}$ represents the full 4D matrix with $V^{(2)}_{n_{i_1}\p n_{i_1}n_{i_2}\p n_{i_2}}$ as elements and $\bs{S}_{1,2}=\bs{S}\otimes\bs{S}$. It is clear that $i_1,i_2$ are not separable but we can do each pair $\{x_1,x_2\},\{y_1,y_2\},\{z_1,z_2\}$ separately.

\section{Calculating Energies}\label{sec.calcenergies}

In order to calculate eigenvalues efficiently, it is important to perform matrix-vector products sequentially. In order to simplify the notation, we restrict this attention to the one-electron problem and define $\bs{B}_j^{(t)}=\bs{S}^{-1/2}\bs{O}^{(t)}_x\bs{S}^{-1/2}$ or $\bs{O}^{(t)}_x\bs{S}^{-1}$ such that the Hamiltonian is now
\begin{equation}
H_{\vec{n}\p n}= \sum_{t}^T\prod_{i=x,y,z}B^{(t)}_{n_i\p n_i}
\end{equation}
where the first summation is over all the terms and $t$ is an arbitrary labelling of the term.
The full matrix vector product for one coefficient $\vec{u}_{\vec{n}\p}$ corresponding to indices $\vec{n}\p=[n_x\p,n_y\p,n_z\p]$ is given as
\begin{eqnarray}\label{eq.naive}
\vec{w}_{\vec{n}\p}&=&\sum_{n_x} \sum_{n_y} \sum_{n_z} B^{(t)}_{n_x\p n_x} B^{(t)}_{n_y\p n_y} B^{(t)}_{n_z\p n_z}\vec{u}_{\vec{n}},\\\label{eq.sequential}
&=&\sum_{n_x}  B^{(t)}_{n_x\p n_x}\left[ \sum_{n_y} B^{(t)}_{n_y\p n_y}  \left[\sum_{n_z}B^{(t)}_{n_z\p n_z}\vec{u}_{\vec{n}}\right]\right],
\end{eqnarray}
with the sum performed over all $L$ values on the grid for each of $x,y,z$. The total summation performed naively using \Eq{eq.naive} requires a scaling of $L^6$ since, for each of the $L^3$ components of $\bs{w}$, one must sums over all $L^3$ components of $\bs{u}$. One can instead perform matrix-vector products sequentially using \Eq{eq.sequential}. This is performed by using intermediate vectors $\vec{\bs{u}}\p$ and $\vec{\bs{u}}^{\prime\prime}$. First, $\vec{\bs{u}}\p$ is formed by performing the sum over $z$ first such that,
\begin{equation}\label{eq.step1}
\vec{u}_{n_x,n_y,n_z\p}\p=\sum_{n_z}B^{(t)}_{n_z\p n_z}\vec{u}_{\vec{n}}
\end{equation}
which involves a sum over $L$ terms for $L^3$ indices. Then $\vec{\bs{u}}^{\prime\prime}$ is  generated by taking $\vec{\bs{u}}\p$ as an input for the sum over $y$ such that,
\begin{equation}\label{eq.step2}
\vec{u}^{\prime \prime}_{n_x,n_y\p,n_z\p}=\sum_{n_y}B^{(t)}_{n_y\p n_y}\vec{u}\p_{n_x,n_y,n_z\p}
\end{equation}
which involves another sum over $L$ terms for each $L^3$ indices. The sum is then performed over $x$ using $\vec{\bs{u}}^{\prime\prime}$ as an input which results in 
\begin{equation}\label{eq.step3}
    \vec{u}_{\vec{n}\p}=\sum_{n_x}B^{(t)}_{n_x\p n_x}\vec{u}^{\prime \prime}_{n_x,n_y\p,n_z\p}.
\end{equation}

Two advantages can be noted.  First,
the three step process of \Eq{eq.step1}-\Eq{eq.step3} requires $3L^4$ operations compared to $L^6$ for \Eq{eq.naive} with the resulting vectors $\vec{u}_{\vec{n}\p}$ being equivalent. This is the computational benefit of having a sum-of-products Hamiltonian. 
The other key benefit is memory usage. One only needs to store a number of $L\times L$ matrices compared to the full Hamiltonian of $L^3 \times L^3$ size. 

For the two-electron problem, the matrices are not separable for $j_1,j_2\in\{x,y,z\}$. This means that the sum that needs to be performed is
\begin{equation}
\vec{w}_{\vec{n}\p}=\sum_{n_{x_1}}\sum_{n_{x_2}}V_{n_{x_1}\p n_{x_1}n_{x_2}\p n_{x_2}}\sum_{n_{y_1}}\sum_{n_{y_2}}V_{n_{y_1}\p n_{y_1}n_{y_2}\p n_{y_2}}\sum_{n_{z_1}}\sum_{n_{z_2}}V_{n_{z_1}\p n_{z_1}n_{z_2}\p n_z{_2}}\vec{u}_{\vec{n}}
\end{equation}
where $V_{n_{j_1}\p n_{j_1}n_{j_2}\p n_{j_2}}$ are the $L^2\times L^2$ matrices for the two-electron Coulomb operator. Performing this matrix vector product sequentially requires $3\times L^6\times L^2=3L^8$ operations. If the Coulomb term was not separated into $3$ products, the total cost of the matrix-vector product would be $L^{12}$.  Storing these $L^2\times L^2$ matrices is the major RAM requirement of these calculations since a fairly large grid is required to converge eigenvalues. Storing the full matrix would not be feasible.

The direct product basis for a two-electron problem has $L^{6}$ components but even this becomes intractable quickly. Therefore, instead of using the full direct product grid, we only include basis functions that have significant overlap with the desired wavefunction(s).  We call this set of basis function labels, $\vec n$, as $\beta$. The size of $\beta$ is the number of one-electron basis functions denoted $M$. Using a general pruned basis $\beta$ complicates the matrix-vector products significantly. The matrix representation is now
\begin{equation}\label{eq.eigform1c}
\left(\sum_{t} \bs{P}_{\beta}^T\left(\bs{B}^{(t)}_x \otimes \bs{B}^{(t)}_y \otimes \bs{B}^{(t)}_z \right)\bs{P}_{\beta}\right)\bs{U}=\bs{U}\bs{E},
\end{equation}
where $\bs{P}_{\beta}$ is a rectangular matrix that projects out the appropriate matrix elements for basis $\beta$.
The sequential summation is now,
\begin{equation}\label{eq.mvherm}
\vec{w}_{\vec{n}\p}=\sum_{n_x}  B^{(t)}_{n_x\p n_x} \sum_{n_y(n_x)} B^{(t)}_{n_y\p n_y} \sum_{n_z(n_x,n_y)}B^{(t)}_{n_z\p n_z}\vec{u}_{\vec{n}}
\end{equation}
where $n_z(n_x,n_y)$ denotes that the summation for dimension $z$ is performed only over those values of $n_z$ that are in the set $\beta$ that have corresponding $n_y,n_z$ values. This is the effect of the projector $\bs{P}_{\beta}$. 
The second summation is performed over all values of $n_y(n_x)$ which indicates that only $n_y$ values that have a corresponding $n_x$ are included while $n_z$ can be any value in $\beta$. The final summation is performed over all $n_x$ in $\beta$ but only generate the output for $\vec{n}\p \in \beta$ which is the effect of the projector $\bs{P}^T_{\beta}$. 
This means that in order to perform matrix-vector products sequentially, intermediate vectors $\vec{\bs{u}}^{\prime },\vec{\bs{u}}^{\prime \prime}$ will have a larger size than $M$. 

If one uses the form of \Eq{eq.mvherm} as stated, then the intermediate vectors $\vec{\bs{u}}^{\prime },\vec{\bs{u}}^{\prime \prime}$ will most likely be larger than necessary. One only needs to retain any intermediate basis functions that includes the union of set $\left\{ n_x,n_y,\bar{n_z} \right\}$ and $\left\{\bar{n_x},n_y,n_z \right\}$, with the $\bar{n_i}$ signifying that all values of $n_i$ are taken while only the combinations $n_j,n_k$ in the retained basis $\beta$ are included. Therefore, the intermediate vectors will still be greater than $M$ but smaller than what would result using \Eq{eq.mvherm}. This is explained well in \Ref{Wodraszka2017} for any number of dimensions and is implemented here.

There is also an approximation that can be made such that intermediate vectors are always the same size. This is the product approximation where the matrix-representation of \Eq{eq.eigform1} is replaced by,
\begin{equation}\label{eq.eigform1p}
\left(\sum_{t} \bs{F}^{(t)}_x \bs{F}^{(t)}_y \bs{F}^{(t)}_z \right)\bs{U}=\bs{U}\bs{E}
\end{equation}
where $\bs{F}_x=\bs{P}_{\beta}^T\left(\bs{S}^{-1/2}\bs{O}^{(t)}_x\bs{S}^{-1/2}\otimes \bs{I}_y \otimes \bs{I}_z \right)\bs{P}_{\beta}$, $\bs{F}_y=\bs{P}_{\beta}^T\left(\bs{I}_x \otimes \bs{S}^{-1/2}\bs{O}^{(t)}_y\bs{S}^{-1/2}\otimes \bs{I}_z\right)\bs{P}_{\beta}$, and
$\bs{F}_z=\bs{P}_{\beta}^T\left(\bs{I}_x \otimes \bs{I}_y \otimes\bs{S}^{-1/2}\bs{O}^{(t)}_z\bs{S}^{-1/2}\right)\bs{P}_{\beta}$ where $\bs{I}_i$ is the identity operator for coordinate $i$.  Performing matrix-vector products is now
\begin{equation}\label{eq.mvprod}
\vec{w}_{\vec{n}\p}=\sum_{n_x(n_y\p,n_z\p)}  B^{(t)}_{n_x\p n_x} \sum_{n_y(n_x,n_z\p)} B^{(t)}_{n_y\p n_y} \sum_{n_z(n_x,n_y)}B^{(t)}_{n_z\p n_z}\vec{u}_{\vec{n}}.
\end{equation}

The main disadvantage of the product approximation is that the Hamiltonian representation is no longer Hermitian.  To justify this claim, consider an approximate sum-of-products approximation where $H_{trunc.}=\sum_i h_{i}h_{i+1}$ with Hermitian conjugate $H_{trunc.}^\dag \approx \sum_j h_{i+1}h_i$.  The action of the operator and its dual are only the same when $h_ih_{i+1}=h_{i+1} h_i$.  
This can be rectified by taking the transpose of the original ordering of the above matrices such that $\bs{F}_x\bs{F}_y\bs{F}_z$  becomes $\frac{1}{2}(\bs{F}_x\bs{F}_y\bs{F}_z+\bs{F}_z\bs{F}_y\bs{F}_x)$. See Ref.~\cite{Cooper2009} for the first numerical application to molecular physics. However, in our context, this approach doubles the computational cost and as will be shown later in Section \ref{sec.oneelectron}, is not necessary for accurate calculations to be made. One simply needs to use Arnoldi iterations as opposed to Lanczos iterations to calculate eigenvalues/eigenvectors \cite{Lehoucq1998b}. This does increase memory requirements as a set of basis vectors need to be stored but, for the calculations performed here, the main memory costs are from storing the Hamiltonian matrix elements for the two-electron Coulomb terms.

\subsection{Choosing basis functions}\label{ssec.basis}
It is impossible to know \emph{a priori} the overlap of the basis functions $d_n(x)$ with the desired eigenfunctions. Choosing functions that are centered in the classically allowed region can provide semi-quantitative accuracy,\cite{Halverson2012} however this is difficult to do in multiple dimensions. A better method is to iteratively improve the basis function by starting with a small basis (motivated by classical phase-space energies) and progressively add more functions around the most important basis functions to improve the description of the wavefunction.\cite{Pipek2012,Shimshovitz2014,Brown2015} In this study, the functions included in set $\beta$ are determined by considering the diagonal elements of the density matrix for the eigenstate.

Consider first the case where we are interested in the ground state and we will generalize to cases where we consider more than one state subsequently.  At the start of the computation, $L$ is fixed and $\beta$ is a subset of $M$ functions chosen from the full $L^3$ basis functions.  After each computation of an approximate wave function, we have $v_{\vec{n}i}$ as the coefficient for basis function $\alpha_{\vec{n}}$.  The importance function of each basis function is then defined by $\mathcal{P}_{\vec{n}}=v_{\vec{n}i}^2$.  If one is interested in optimizing the pruned basis for more than one eigenstate at a time, then the total importance function is the sum of the importance vector for each targeted state.

The importance vector determines which basis function are pruned and determines which basis functions will be added for the next iteration.  Those basis function with importance values below a cutoff threshold are removed and not considered in future iterations.  Expansion of the set $\beta$ occurs near the functions with the largest importance values.
Additions are made in all $3\times4$ directions in phase-space: three for each dimension $x,y,z$ and four for each choice of $\pm m,\pm k$. Special care needs to be (and was) taken to exclude redundant function that are already in $\beta$ and to exclude $\inm \in 2\mathbb{Z}+1,k=0$.  For basis functions of $m=l,k=0$ the additions to the set $\beta$ are only $m=l\pm2,k=0$, $m=l\mp1,k=1$ and $m=l,k=1$.

For all calculations in this paper, we are using basis functions in $x,y,z$ that can be symmetry adapted with respect to inversion of single coordinate e.g. $f(-x,y,z)=\pm f(x,y,z)$. All systems studied in this paper have symmetry about the origin in at least two coordinates and therefore fewer basis functions will be needed to represent their wavefunction. This is true even though the full wavefunction (including correlation) does not respect this single-coordinate inversion symmetry. The asymmetry introduced by correlation is a much smaller portion of the total wavefunction and therefore requires fewer anti-symmetric basis functions to describe it accurately. This means that at the origin, $k\in 2\mathbb{Z}$ are symmetric functions while $k\in 2\mathbb{Z}+1$ are anti-symmetric functions.  All other functions are localized in four positions in phase-space ($\pm m\dx,\pm k \pi/\dx$) except for $k=0$ which is localized at two points. The indices are then relabeled such that positive $m$ means moving outward from the origin in symmetric functions while a decrease in negative $m$ from means moving outward from the origin in anti-symmetric functions.

\section{One-Electron calculations}\label{sec.oneelectron}
There are four parameters that one controls, the lattice spacing in position space $\dx$, the number (and type) of quadrature points used for the Coulomb terms, the number of underlying $d_n\left(x\right)$ functions, and whether the Hermitian (\Eq{eq.mvherm}) or product approximated (\Eq{eq.mvprod}) Hamiltonian representation is used. The hydrogen atom is used as the test for all parameters. The basis is symmetry adapted for $x,y,z$, with basis functions located at $...,-2\dx,-\dx,0,\dx,2\dx,...$ and the hydrogen nucleus located at $0,0,0$. The size of the position grid is always of the size $L_k=4\mathbb{Z}-1$ so that $m=0,k=0$ is in the basis with with an odd number of $m$ values on either side. The symmetric orthogonalized form of \Eq{eq.eigform1} with Hermitian matrix-vector products is used unless otherwise stated. Also, the underlying $d_n\left(x\right)$ grid is only has large as necessary except for section \ref{sec.bigvsmall}. Implicitly restarted Arnoldi via ARPACK\cite{Lehoucq1998b} is used to calculate eigenvalues/eigenvectors. When performing test of grid spacing and when comparing the Hermitian and product approximations, $19$ Gauss-Jacobi quadrature points were used.  However, 13 quadrature points were used for all multi-electron calculations consistent with our findings in the next subsection.

\subsection{Convergence of quadrature approximation}
\begin{figure}[ht]
\caption{Convergence of ground state eigenvalue of hydrogen atom using a grid spacing of $\sqrt{\pi}$ using an increasing number of Gauss-Jacobi quadrature points. \label{fig.jacconv1s}}
\centering
\includegraphics[width=0.5\textwidth]{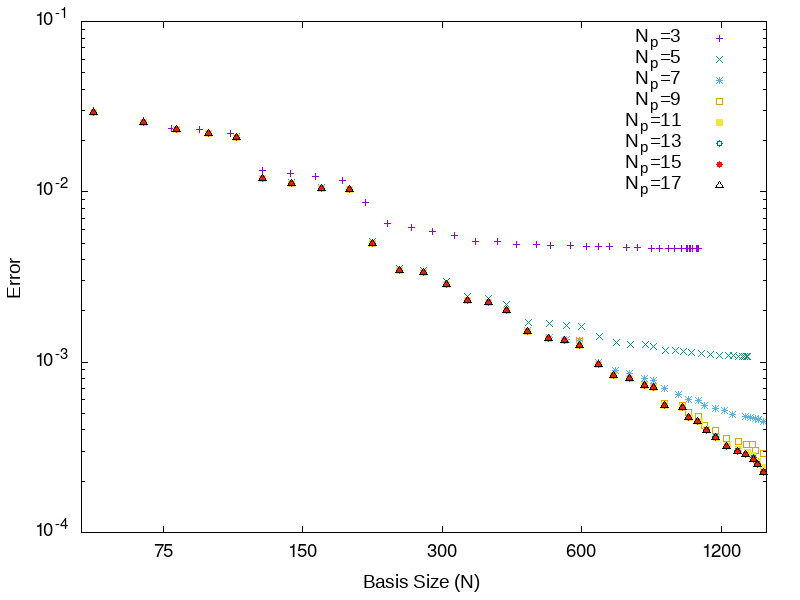}
\end{figure}

\begin{figure}[ht]
\caption{Convergence of 2P state eigenvalue of hydrogen atom using a grid spacing of $2\sqrt{\pi}$ using an increasing number of Gauss-Jacobi quadrature points. \label{fig.jacconv2p}}
\centering
\includegraphics[width=0.5\textwidth]{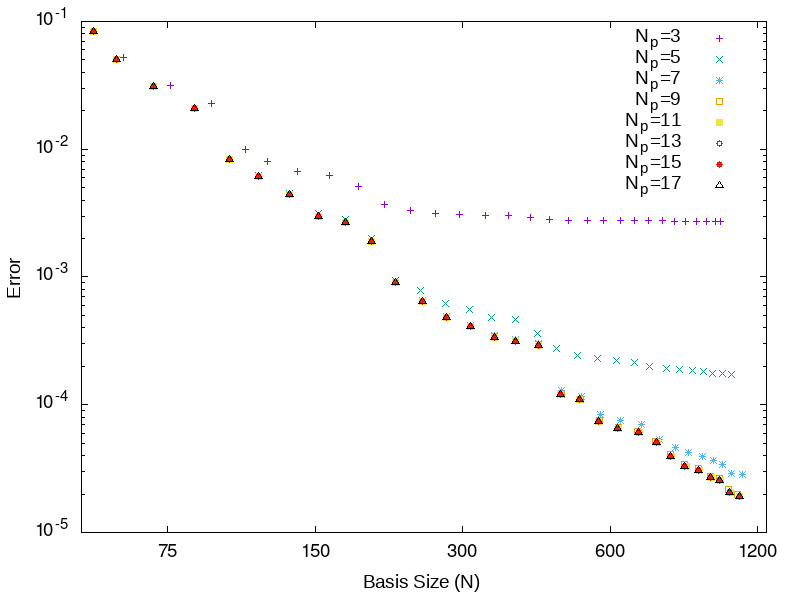}
\end{figure}

The convergence of the Gauss-Jacobi quadrature approximation can be seen in \Fig{fig.jacconv1s} and \Fig{fig.jacconv2p} for the 1S (with $\dx=\sqrt{\pi}$) and the 2P state (with $\dx=\sqrt{\pi}$) respectively. For the 1S state with  $\dx=\sqrt{\pi}$, chemical accuracy $\epsilon<10^{-3}$ Hartree can be achieved with only $7$ quadrature points. This is not true if one uses $\dx=2\sqrt{\pi}$ where $11$ quadrature points are required. The reason for this is that the maximum momentum index required for chemical accuracy is $k=15$ with $\dx=2\sqrt{\pi}$ while only $k=7$ for $\dx=\sqrt{\pi}$. A more highly oscillatory basis requires a higher degree polynomial to represent and therefore more quadrature points.  
 Therefore, all later calculations are performed with 13 Gauss-Jacobi quadrature points which is more than enough to achieve chemical accuracy without the quadrature error entering the calculation.
One can also use Gauss-Legendre quadrature but the number of points required to achieve chemical accuracy is generally larger by about $50\%$.

\subsection{Hermitian vs Product Approximation}
\begin{figure}[ht]
\caption{Convergence of 1S state of a hydrogen atom using a grid spacing of $\sqrt{\pi}$ using the Hermitian and Product approximation. \label{fig.hermvprod}}
\centering
\includegraphics[width=0.5\textwidth]{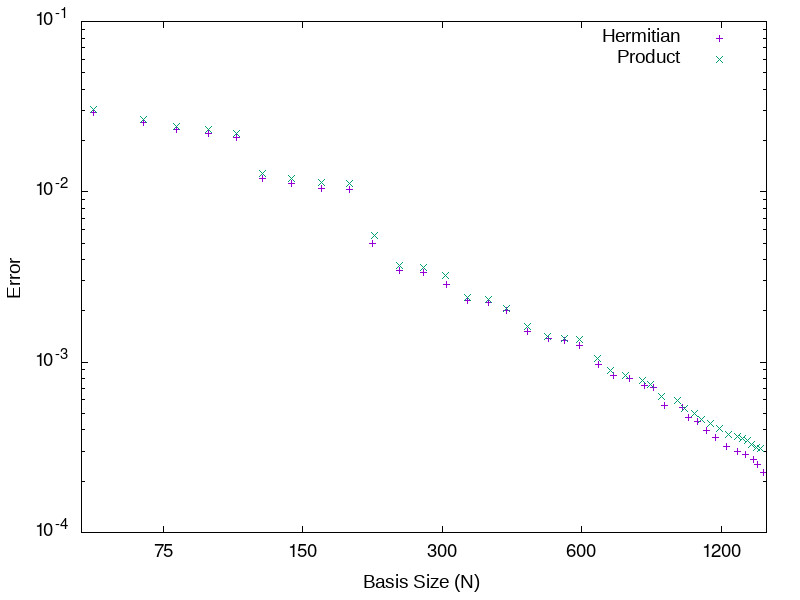}
\end{figure}

For the 1S state of hydrogen with a position spacing $\dx=\sqrt{\pi}$, the comparison between the Hermitian (of \Eq{eq.mvherm}) and product approximated form (of \Eq{eq.mvprod}) is shown in \Fig{fig.hermvprod}. It is clear that the product approximation only manifests when high accuracy is required. For this calculation, that occurs after chemical accuracy has been achieved. This is not the case if $\dx=2\sqrt{\pi}$ for which the difference is evident when the error is approximately $0.002$ $E_h$ and error below $0.001$ $E_h$ requires a larger basis. Even with this disadvantage, the speed of the calculation using the product representation is much faster as intermediate basis sizes can be upwards of six times larger than the retained basis.

\subsection{Using different sizes of modulated Gaussians phase-space localized functions\label{sec.bigvsmall}}
\begin{figure}[ht]
\caption{Convergence of 1S state of a hydrogen atom using a grid spacing of $\sqrt{\pi}$ using a Large or Small underlying grid \label{fig.largevsmall}}
\centering
\includegraphics[width=0.5\textwidth]{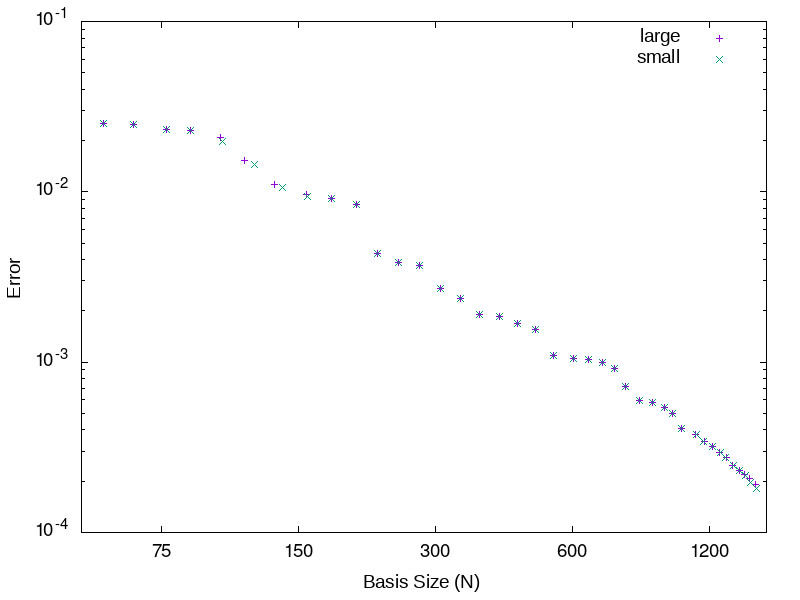}
\end{figure}

We have also testing using different sizes of modulated Gaussian (\Eq{eq.cdmk}) grids for the 1S state of a hydrogen atom. The large grid includes indices $m\in\left[-21,21\right],k\in \left[0,22\right]$. The small grid has indices $m\in\left[-7,7\right],k\in \left[0,13\right]$. The basis functions after the 40 expansion iterations shown have indices $m\leq 6$ and $k\leq 11$. In \Fig{fig.largevsmall}, the difference between the large and small grid of $d_n\left(x\right)$ functions are insignificant near convergence. In fact, the coefficients of the eigenvector are the same to several (approximately four) digits for a given $n$. Knowing this, we only use as many functions as is necessary to cover a large enough region of phase-space to converge the calculations. This is especially important for the two-electron calculations when memory resources become the constraining factor.

\subsection{Convergence 1S, 2S, and 2P hydrogen energy levels using different grid spacing}
\label{sec:wilson_ convergence}
 The S orbitals are determined with only even basis functions for $x,y,z$ while the 2P state is determined by using even basis functions for $x,y$ and odd basis functions for $z$. Without symmetry adaption, the number of basis functions would be $8=2^3$ times larger.  

\begin{figure}[ht]
\caption{Convergence 1S, 2S, and 2P states of hydrogen with a grid spacing of $\sqrt{\pi}$. The order of convergence is fit for each state.\label{fig.hydconv1p0}}
\centering
\includegraphics[width=0.5\textwidth]{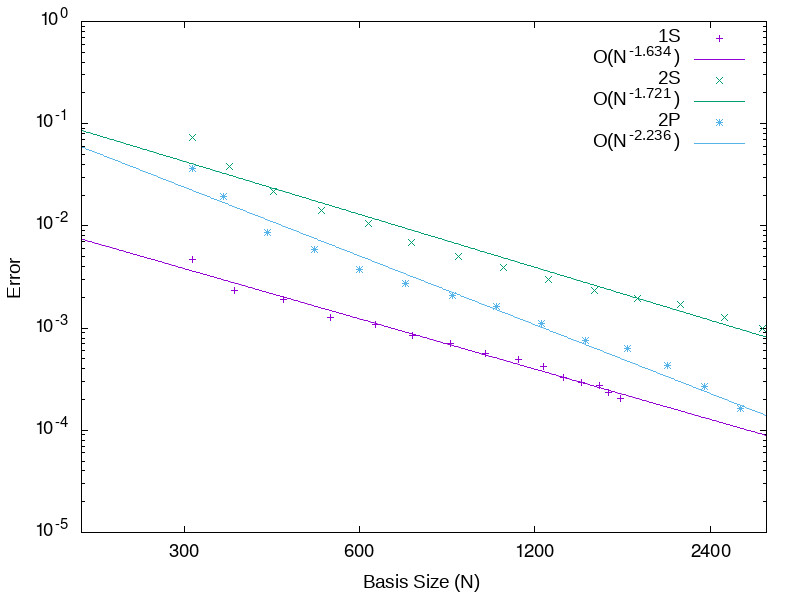}
\end{figure}

The convergence of hydrogen energy levels is determined by growing the basis using only the wavefunction from the desired state (see Subsection \ref{ssec.basis}). The convergence has been examined for the spacing of $\dx=\sqrt{\pi}$ and $\dx=2\sqrt{\pi}$ with each state being optimized independently. The discontinuous derivative of the 1S and 2S states results in a slower convergence than for the 2P state.  For $\dx=\sqrt{\pi}$ (shown in \Fig{fig.hydconv1p0}), the convergence of the 1S, 2S and 2P states is $N^{1.63}$, $N^{1.72}$ and $N^{2.236}$ respectively. When $\dx=2\sqrt{\pi}$, the $2S$ and $2P$ states converge much more quickly ($N^{2.602}$ and $N^{5.563}$ respectively) compared to $1S$ only having a convergence of $N^{0.901}$. This is why multiresolution wavelet PSL basis sets are desirable \cite{White2017}. Unfortunately, the Wilson basis does not have this property but we return to this point in Section \ref{ssec.gausslets}.

\begin{figure}[ht]
\caption{Convergence 1S, 2S, and 2P states of hydrogen with a grid spacing of $2\sqrt{\pi}$. The order of convergence is fit for each state.\label{fig.hydconv2p0}}
\centering
\includegraphics[width=0.5\textwidth]{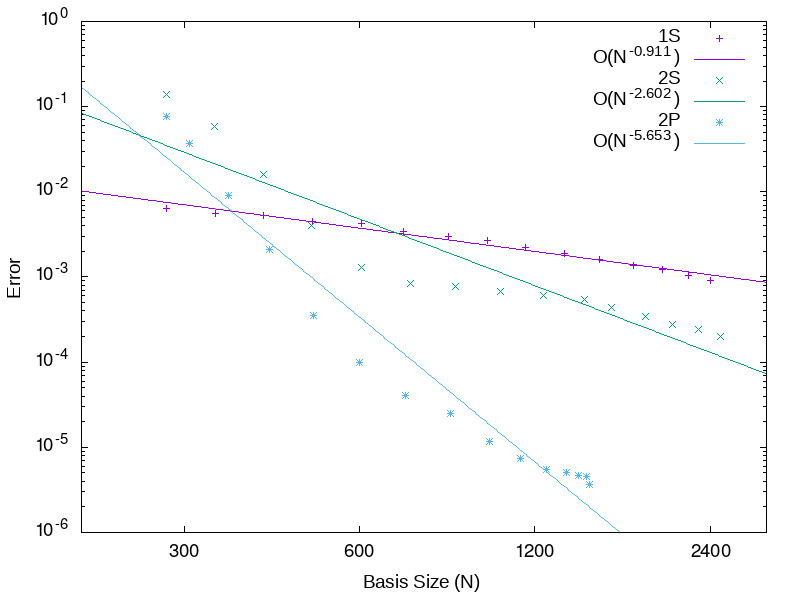}
\end{figure}

\newpage
\section{Two-Electron calculations}\label{sec.twoelectron}
To perform two-electron calculations, we simply take the set of one electron basis functions and make a direct-product to add the other three dimensions. 
\begin{equation}\label{eq.6dbasis}
b_{\vec{n}_1\vec{n}_2}=b_{\vec{n}_1}\otimes b_{\vec{n}_2}
\end{equation}
where $\vec{n}_1=[n_1,n_2,n_3]$ and $\vec{n}_2=[n_4,n_5,n_6]$ are both contained in the pruned basis function set $\beta$. This means that no product approximation is made in the individual two-electron Coulomb terms for $x,y,z$ regardless of whether one uses \Eq{eq.mvherm} or \Eq{eq.mvprod}. The benefit of using \Eq{eq.6dbasis} is that spin can be taken into account if desired. 

For example, one can differentiate between singlet and triplet states by enforcing symmetry under exchange. This is explicitly written as,
\begin{equation}
b^{(s)}_{\vec{n}_{1},\vec{n}_{2}}=(b_{\vec{n}_{1},\vec{n}_{2}}+b_{\vec{n}_{2},\vec{n}_{1}})/2
\end{equation}
for the singlet state and
\begin{equation}
b^{(t)}_{\vec{n}_{1},\vec{n}_{2}}=(b_{\vec{n}_{1},\vec{n}_{2}}-b_{\vec{n}_{2},\vec{n}_{1}})/2
\end{equation}
for the triplet state. Even when using the product approximation (and Arnoldi iterations), symmetry adaptation isolates singlet and triplet states effectively as the product approximation error is small.

If one uses \Eq{eq.mvherm} then singlet/triplet labels can be obtained with a single matrix-vector product for each Lanczos iteration using symmetry adapted Lanczos.\cite{Wang2001} This is because the symmetry projection operator $\bs{P}$ commutes with the Hamiltonian such that $\bs{PHv}=\bs{HPv}$. Therefore, a sum of vectors with different symmetries applied to the Hamiltonian, followed by the symmetries then being projected out is equivalent to applying the Hamiltonian to each vector with distinct symmetry separately. One then stores, and calculates eigenvalues from, separate Lanczos tridiagonal matrices for each symmetry. Note that symmetry adapted Lanczos was not implemented in this pilot study.

\begin{figure}[ht]
\caption{Convergence of $^1S$ ground state of the helium atom with a grid spacing of $0.45\sqrt{\pi}$. The order of convergence is fit to be $O(M^{-0.726})$ for $\bs{S}^{-1/2}\bs{HS}^{-1/2}$ and $O(M^{-0.781})$ for $\bs{HS}^{-1}$ where $M=N^2$.\label{fig.helium1s}}
\centering
\includegraphics[width=0.9\textwidth]{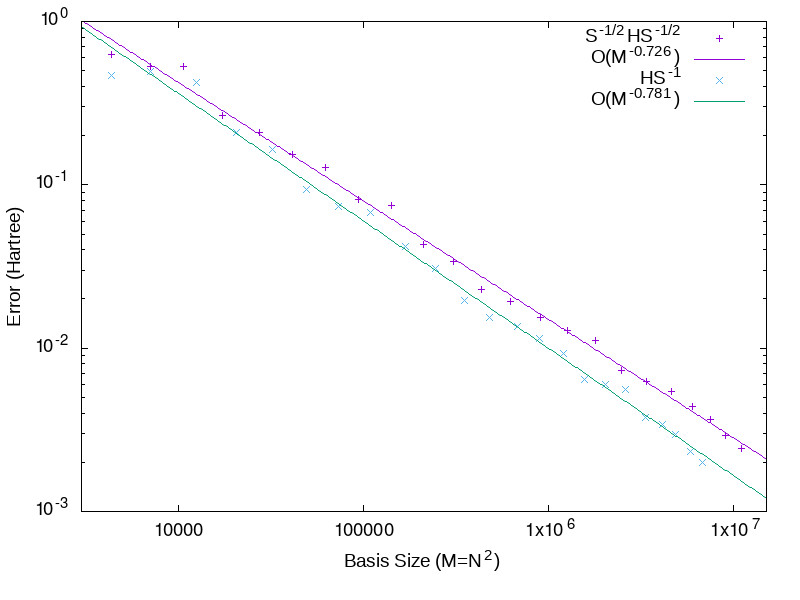}
\end{figure}

The first test case is for the $^1S$ ground state of the helium atom. The convergence of the basis with  $\dx=0.45\sqrt{\pi}$ on a grid of $L_k=11,L_m=11$ is found to be $M^{-0.726}$ (where $M=N^2$) for the symmetric basis and $N^{-0.781}$ for the biorthogonal basis as shown in \Fig{fig.helium1s}. The most accurate energy calculated is found to be $-2.9013$ $E_h$ with the exact solution being $-2.90372...$ $E_h$\cite{Kurokawa2008}. This required a basis of $3,325\times3,325=11,055,625$ functions. This calculation took just under five hours using 16 cores on a Xeon(R) E5-2640 processor and required $28.5$GB of RAM. The majority of the RAM usage ($22$ GBs) were required to store the two-body Coulomb matrix representation at each of the 13 quadrature points. A value of $-2.9017$ $E_h$ using the biorthogonal representation used $2603\times2603=6,775,609$ basis functions and required just under $3$ hours of computing time using $26.2$GB of RAM on the same computer. The reduction of approximately $20\%$ in the one-electron basis size is fairly consistent across all calculations examined when using the biorthogonal representation relative to the orthogonal basis. 

We also performed calculations on the hydrogen molecule at a distance of $0.74$\AA\, and obtained the lowest two singlet states along with the lowest energy triplet state.
The sampling grid was $L_m=11,L_k=11$ with $\dx=1.3\sqrt{\pi}$. Optimizing the ground state separately resulted in an energy of $-1.8872$ $E_h$ with a basis size of $3180\times3180=10,112,400$. 
When optimizing for all three eigenvalues, a basis of $2898\times2898=8,398,404$ obtained energies of $-1.8865$ $E_h$ for $X\,^1\Sigma_g^+$, $-1.4961$ $E_h$ for $b\,^3\Sigma_u^+$ and $-1.4254$ $E_h$ for $B\, ^1\Sigma_u^+$. All three of these values are more converged than those calculated with \Ref{Jerke2018}. This is especially true for $B\, ^1\Sigma_u^+$ which is $0.007E_h$ lower in energy.

\section{Hartree-Fock}\label{sec.HF}
Hartree-Fock or mean-field ansatz is also possible using the Wilson basis functions.  This is because the Coulomb basis is not truncated but rather remains a direct product of basis functions, \Eq{eq.6dbasis}.  The Hartree-Fock equations are derived by taking the gradient of the energy functional of charge density matrices.  The resulting operator is the Fock matrix given in standard notation as \cite{Szabo1996}
\begin{eqnarray}
F_{\vec{n}\p \vec{n} }=H_{\vec{n}\p \vec{n} }+\sum^{N_e/2}_{a}\sum_{\vec{m} \vec{j} }^{\beta} C_{\vec{m}\,a} C_{\vec{j} a}^* [ 2 (\vec{n}\p \vec{n} |\vec{j}\vec{m} ) -(\vec{n}\p \vec{m} |\vec{j}\vec{n} ) ] 
\end{eqnarray}
with $(\vec{n}_1\vec{n}_1\p|\vec{n}_2\vec{n}_2\p)=I_{\vec{n}_1\p\vec{n}_2\p,\vec{n}_1 \vec{n}_2}(|r_1-r_2|^{-1})$.
We can achieve a more efficient algorithm by rearranging the action of the second term as follows
\begin{eqnarray}
&&\sum_{\vec{n}}^\beta\left(\sum^{N_e/2}_{a}\sum_{\vec{m} \vec{j} }^{\beta} C_{\vec{m}\,a} C_{\vec{j} a}^* [ 2 (\vec{n}\p \vec{n} |\vec{j}\vec{m} ) -(\vec{n}\p \vec{m} |\vec{j}\vec{n} ) ]\right)b_{\vec{n}}\label{eq.sum1}\\
&=&\sum_{\vec{n}}^\beta\sum^{N_e/2}_{a}\sum_{\vec{j} }^{\beta} C_{\vec{j} a}^*
\sum_{\vec{m} }^\beta(\vec{n}\p \vec{n} |\vec{j}\vec{n} )
  \left[ 2C_{\vec{m}\,a}  b_{\vec{n}}  -C_{\vec{n}\,a} b_{\vec{m}}  \right]\label{eq.hfint}
\end{eqnarray}
This second form allows us to perform the action of the Fock operator without constructing the matrix. 
In some cases, the first form could be more efficent if the number of electrons is very high but the second form is what has been implemented.  The integral driven procedure scales as $3N_e(N^2+N^{8/3})/2$.  If instead one first forms the charge density matrix, $P_{\lambda \sigma}=\sum_a^{N_e/2}C_{\lambda a} C_{\sigma a}$, then performs the remaining summations in \Eq{eq.sum1}, a cost of $(N^2)(3N^{4/3})/2$ is expected.  It is therefore cheaper to use the density matrix approach only when the number of electrons is less than $N^{2/3}$.
For all calculations performed here, $N_e\ll N^{2/3}$, so the integral driven approach was utilized.

\begin{figure}[ht]
\caption{Convergence of $^1S$ ground state of the helium atom using Hartree-Fock with a grid spacing of $0.45\sqrt{\pi}$. The order of convergence is fit to be $O(N^{-1.561})$.\label{fig.helium1sHF}}
\centering
\includegraphics[width=0.9\textwidth]{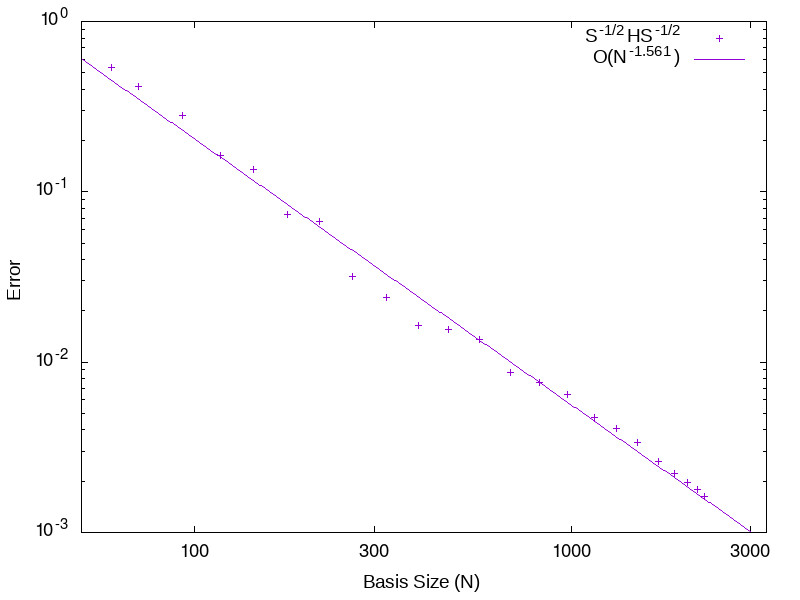}
\end{figure}

The convergence of the Hartree-Fock calculation of the helium atom is shown in \Fig{fig.helium1sHF}. The convergence using the symmetric basis is $N^{-1.561}$ which is essentially the same convergence as the full two electron calculation, but $N$ is denoted by the size of the one-electron basis here. The most accurate value obtained is $-2.86012$ $E_h$ using a basis of $2525$ functions. The exact value is $-2.8618$ for an error of less than $0.002$ $E_h$.

\subsection{Four electron system using Hartree-Fock}
Using \Eq{eq.hfint}, we study the four-electron system LiH at a spacing of $3$ $a_0$ with $H$ at $-0.75 \,a_0$ and Li at 2.25$a_0$. The system geometry is such that both atoms are on the $x$-axis.  Consequently, only even-parity functions are necessary for $y$ and $z$ directions reducing the necessary basis size by a factor of four. 
By choosing the step size in each direction the Couloumb matricies are the same in each direction.  By performing the same transform to the symmetry adapted basis to each coordinates maintains the equivalence of the Coulomb matrices in each direction.  Hence, we can store a single $L\times L \times L \times L$ Couloumb matrix at each quadrature point rather than maintaining separate representations in each Cartesian direction.

The basis functions for this calculation were sampled from a grid of $L_m=15$, $L_k=7$ with $N_p=13$ quadrature points and required $\approx 17$GB of memory.  The majority of this memory overhead comes from the thirteen $(15\times7)^4$ two-electron matrices representing approximately 13 GB of memory.

To test the performance of the Wilson basis on the LiH system, we compared against the standard basis sets of STO-3G and 3-21G with respective Hartree-Fock energies of $-7.8623$ and $-7.9295$ respectively. Of the total $L^3=1,157,625$ Wilson basis functions, a pruned basis set $\beta$ of size $N=2048$ was necessary to achieve STO-3G accuracy and $N=3511$ was needed to reach 3-21G accuracy.

We performed the calculations using 16 cores of a Xeon(R) E5-2640 processor with 64GB RAM, part of Dartmouth's Discovery Cluster. To obtain STO-3G accuracy took 10.3 hours while obtaining 3-21G accuracy took approxmately $38$ hours. 

\section{Extensions of the Wilson basis set}\label{sec.tests}

The necessity of describing both diffuse and tightly bound eigenfunctions simultaneously complicates the usage of Wilson basis functions. Therefore, the Wilson basis will probably be more successful when combined with pseudopotentials. Instead of using pseudopotentials, two alternative ideas are tested.  

The first is to utilize a multiresolution basis and the second is to augment the Wilson basis with projected Slater type orbitals.  The multiresolution
gausslets\cite{White2017} are more effective at describing both diffuse and tightly bound eigenfunctions but, as we point out in the next subsection, there are cases where the Wilson basis remains better.

The second method makes use of the sum-of-products form of the STO-$n$G basis. We showed that accurate energies of one electron in a  LiH$_2$ potential can be obtained by replacing high momentum functions with the STO-$6$G basis centered on the Li nuclei. This second technique will likely be most useful when using an expansion of the eigenfunction as a sum-of-products described in \Ref{Thomas2018} and \Ref{Jerke2018}. 

We will close this section with a tour through other ideas for improving the overall numerical implementation.  The final subsection will highlight several paths for improvement that were not done in this study.

\subsection{Using gausslets}\label{ssec.gausslets}
Earlier in section \ref{sec:wilson_ convergence}, we noted that the convergence rates of the 1S state is faster when the the Wilson function grid spacing is small but the 2S and 2P states converge faster with larger grid spacing.  This suggests that a single choice of resolution will not work for multiple states.  In this subsection, we explore multiresolution approaches and compare them against the Wilson basis.  We find that the convergence rates of the Wilson basis are faster for individual states but the multiresolution functions can better capture simultaneous convergence of multiple states.

It has been shown\cite{Shimshovitz2012b} that using multiresolution wavelets for Coulomb potentials can provide a significant reduction in the number of basis functions required for one-dimensional convergence. However, most wavelets used previously are only defined on a grid, and as the electron-nuclei and electron-electron potentials are unbounded, the Hamiltonian representations become non-trivial when performing quadrature due to the singulaity at various points in space. One then either needs to use a pseudo-potential\cite{Genovese2008} or soften the discontinuity\cite{Shimshovitz2012b}. 

In another approach to this problem, White\cite{White2017} developed a wavelet theory using ``gausslets'' that are defined as certain linear combinations of a equally spaced grid of Gaussians. 

This allows a multi-resolution wavelet transform to be performed on a grid of Gaussian functions.  Starting from a single wavelet, a set of three transformations can be used to decrease the resolution systematically: a transformation to an even function, a transformation to an odd, and a scaling transform.  The even/odd symmetries are defined by inversion about the center of the gausslet.  Here we consider a calculation for the electronic energy of the hydrogen atom.  

The grid of Gaussians is defined as
\begin{equation}
g_i\left(x\right)=\sqrt{\delta x}\exp\left[-\frac{1}{2}\left(i-3\delta x \,x \right)^2\right]
\end{equation}
with $\delta x$ between each Gaussian on the grid.  For each cartesisan direction, we consider a one-dimensional grid with $i$ ranging between $-4208$ and $4208$.  The overlap integral for two Gaussians on the grid is
\begin{equation}
S^{(g)}_{i\p i}=\frac{\sqrt{\pi}}{3}\exp\left[-\frac{1}{4}\left(i-i\p\right)^2\right].    
\end{equation}
The kinetic energy operator integral is
\begin{equation}\label{eq.gkin}
I_{i\p i}^{(g)}(T)=- \frac{1}{8\delta x^2}\left(-2+(i-i\p)^2\right)S^{(g)}_{i\p i}, 
\end{equation}
and the one-electron Coulomb integral for each Legendre quadrature point $p_{\alpha}$ is
\begin{equation}\label{eq.gcou}
V^{(g)}_{i\p i}(p_{\alpha},R_i)=\exp\left[-\frac{1}{16}(i+i\p-6 \delta x R_i)^2 (1+p_{\alpha})^2\right]S^{(g)}_{i\p i}.
\end{equation}
The full Coulomb operator is 
\begin{equation}
I_{\vec{n}\p\vec{n}}\left(\frac{1}{r-r\p}\right)\approx \frac{\sqrt{\pi}}{\delta x}\sum_{\alpha=1}^{N_p} w_{\alpha}V^{(g)}_{n_x\p,n_x}(p_{\alpha},R_x)V^{(g)}_{n_y\p,n_y}(p_{\alpha},R_y)V^{(g)}_{n_z\p,n_z}(p_{\alpha},R_z)   
\end{equation}

We use the $\mathcal{G}_{10}$ gausslet with application of the $W_{652}$ wavelet transform repeated $4$ times and a Gaussian grid spacing of $\delta x=0.015$. This choice was found to give the best results using the methodology described in the present paper. The matrix elements were calculated by generating the transformation matrix $\bs{G}$ for each basis function at position $n=(m,k)$
\begin{equation}\label{eq.wavelets}
    w_{n}(x)=\sum_i G_{ni}g_i(x)
\end{equation}
with $G_{ni}$ representing a vector of coefficients in $\bs{G}$ for $n=(m,k)$ such that $m=-8,-7..,0,...,7,8$ for $k=0$ and $m=-7.5,...-0.5,0.5,...,7.5$ for the even and odd wavelets at $k=1,2,...,8$. The centers of the wavelets change depending on the $k$-dependant spacing of the wavelets $\dx^{(k)}$ given by $\dx^{(7)}=\dx^{(8)}=9\delta x$, $\,\dx^{(5)}=\dx^{(6)}=27\delta x$, $\,\dx^{(3)}=\dx^{(4)}=81\delta x$ and $\,\dx^{(0)}=\dx^{(1)}=\dx^{(6)}=243\delta x$. The even and odd wavelet transforms shift the center by a factor of $\dx_k/2$ and is responsible for the different $m$ indexing between $k\neq 0$ and $k=0$.

We generate the basis by starting with gausslet $\mathcal{G}_{10}$,\cite{White2017} defined by the coefficient vector $\vec{G}^{(10)}$ with non-zero elements from $i=[-68,68]$ and centered at $x=0$. The odd $W_{652}$ wavelet transform is applied to $\vec{G}^{(10)}$ to obtain the $G_{(m=1/2,k=8),i}$ coefficients, the even $W_{652}$ wavelet transform is applied to obtain the $G_{(m=1/2,k=7),i}$ coefficients. We obtain another gausslet denoted  $\vec{G}^{(10,6)}$ also centered at $x=0$ with lower resolution by applying the $W_{652}$ scaling transform to the original $\mathcal{G}_{10}$.  This process is repeated, by applying the odd, even, and scaling transforms $\vec{G}^{(10,6)}$, to  to obtain $G_{(m=1/2,k=6),i}$, $G_{(m=1/2,k=5),i}$ and  $\vec{G}^{(10,6,6)}$ respectively, which are now spaced $27\delta x$ apart. The resolution is further decreased using the same three transforms (odd, even and scaling) to obtain $G_{(m=1/2,k=4),i}$, $G_{(m=1/2,k=3),i}$ and  $\vec{G}^{(10,6,6,6)}$ with spacing $81\delta x$. After the final set of transforms, we obtain $G_{(m=1/2,k=2),i}$,$G_{(m=1/2,k=1),i}$ and $\vec{G}^{(10,6,6,6,6)}$  with spacing $243\dx$. Then $G_{m=0,k=0,i}$ is defined as $\vec{G}^{(10,6,6,6,6)}$. To obtain other $m$ values, one simply has to shift the coefficients by the appropriate number of positions for each resolution level. Namely $G_{(m,k),i}=G_{(m=1/2,k),i+(m-1/2)\times s}$ where $s=9$ for $k=7,8$, $s=27$ for $k=5,6$, $s=81$ for $k=3,4$ and $s=243$ for $k=1,2$. For $k=0$ $G_{(m,k),i}=G_{(0,0),i+m\times 243}$. The number of non-zero elements in each row of $\bs{G}$ is $252, 706, 2094, 4528$ for each level respectively.

The calculation of the matrix elements is then simply a contraction of the full operator matrices as
\begin{equation}
I^{(g)}\left(O\right)=\bs{G}\bs{O}\bs{G}^T
\end{equation}
where $\bs{O}$ is one of $\bs{T^{(g)}}$ or $\bs{C^{(g)}}$ described by \Eq{eq.gkin} and \Eq{eq.gcou} respectively, and $\bs{G}$ has elements $G_{ni}$ as generated from wavelet transforms and translations. Clearly, very large matrices are needed in one dimension to calculate matrix elements in this fashion. However, the matrices are banded due to all matrix elements depending on the banded $S_{i\p i}^{(g)}$ such that the calculation and storage requirements only grow linearly with increasing number of wavelets. 

The other issue is that a larger number of quadrature points are required to calculate accurate energy levels. With the Wilson basis, as few as seven Gauss-Jacobi quadrature points are needed for the quadrature error to be less than 0.001$a_0$. For the gausslet basis, $51$ Gauss-Legendre quadrature points are required to achieve the same precision. If a smaller $\delta x$ spacing is used than even more quadrature points are required. Also, Gauss-Jacobi quadrature does not work for gausslets, most likely due to the discontinuity in $V_{i\p i}^{(g)}$ causing problems here but are not important in the Wilson basis. Another issue is that the basis is not variational with respect to quadrature points. If fewer points than necessary are used, one can obtain energies that are lower in energy than the exact value.

\begin{figure}[ht]
\caption{Convergence 1S, 2S states of hydrogen with an underlying Gaussian grid spacing of $0.015a_0$. The bracketed values indicate which states were optimized simultaneously. One can see that there is only a small decrease in accuracy when optimizing both the 1S and 2S state as opposed to only the 1S state. \label{fig.gausshydconv}}
\centering
\includegraphics[width=0.9\textwidth]{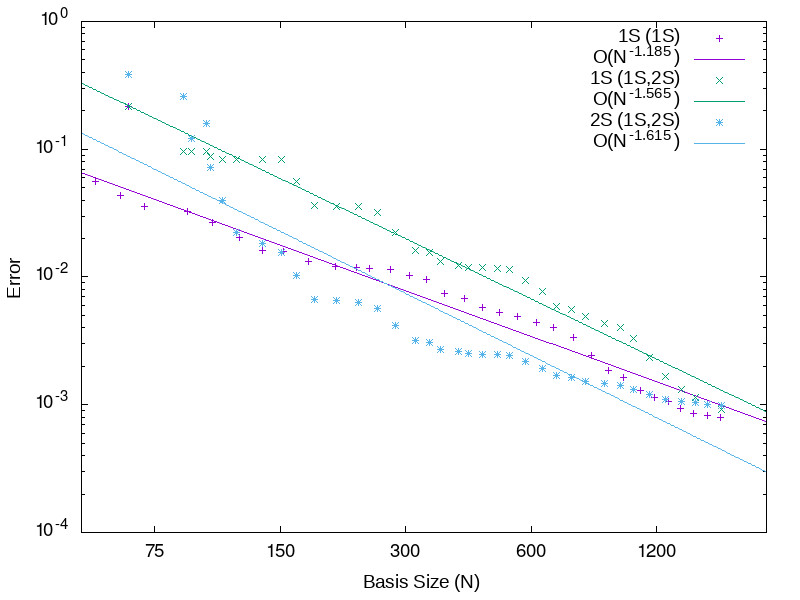}
\end{figure}

The advantage of the gausslet/wavelet basis comes from trying to describe two different eigenfunctions simultaneously. As can be seen from \Fig{fig.gausshydconv}, the accuracy of the 1S state is only marginally worse (for a given basis size) if both the 1S and 2S state are optimized simultaneously, as opposed to only optimizing the 1S state. To obtain chemical accuracy for only the 1S state requires 1366 basis functions, while obtaining chemical accuracy for both the 1S and 2S states requires 1717 basis functions. 

This result is very different from what is found with the Wilson basis. To obtain chemical accuracy only optimizing the 1S state requires only 823 functions while obtaining chemical accuracy of both 1S and 2S states requires 2974 functions. Therefore, using gausslets should allow fewer necessary basis functions for describing multiple states at once. Symmetry adapted wavelets were used in the calculation along with Hermitian matrix-vector products.

\subsection{Augmenting the Wilson basis with Gaussian Slater type orbitals}
Another possible improvement is the ability to combine the Wilson basis with well developed quantum chemistry basis sets. As an example, we will find the ground state of one-electron in a potential of a Li nuclei at the origin and two H nuclei at $\pm3a_0$. The first thing to note is that the STO-$n$G is a basis in sum-of-products form such that
\begin{eqnarray}
\mbox{STO-$n$G}&=&\sum_{i=1}^n c_i\left(\frac{2\alpha_i}{\pi}\right)^{3/4}\exp\left[-\alpha_i (r-r_i)^2\right] \nonumber\\
&=&\sum_{i=1}^n c_i\left(\frac{2\alpha_i}{\pi}\right)^{3/4}\exp\left[-\alpha_i (x-R_x)^2\right]\exp\left[-\alpha_i (y-R_y)^2\right]\exp\left[-\alpha_i (z-R_z)^2\right]
\end{eqnarray}
Next, we can expand each of the $\exp\left[-(x-R_i)^2\right]$ Gaussians in terms of the Wilson basis. 
We wish to describe the portion of phase-space that has contributions from all three nuclei with the Wilson basis set. Therefore, we partition phase-space such that the high momentum region ($k$ greater than cutoff $k_c$) will be described by the projection of the STO-$n$G basis into the $k>k_c$ Wilson basis functions.  Thus, 
for $k>k_c$ the STO-$n$G basis is expanded as
\begin{equation}
\exp\left[-\alpha_i\left(r-r_A\right)^2\right]=\sum_{m,k>k_c}s^{(i)}_{m_x,k_x}p_{m_x,k_x}\left(x\right)s^{(i)}_{m_y,k_y}p_{m_y,k_y}\left(y\right)s^{(i)}_{m_z,k_z}p_{m_z,k_z}\left(z\right)
\end{equation}
where $s^{(i)}_{m_x\p,k_x\p}=\sum_{n_x=(m_x,k_x)} S^{-1/2}_{n\p_x n_x}\int_{-\infty}^{\infty}\exp\{-\alpha_i \left(x-R_i\right)^2\}d_{m_x,k_x}\left(x\right)$ and $p_{m,k}(x)=\sum_{n_x} S^{-1/2}_{n_x n\p_x}d_{m_x\p,k_x\p}\left(x\right)$. These states are then combined with the Wilson basis set for $k<k_c$.

These functions are orthogonal to the $k\leq k_c$ Wilson basis but are not orthogonal to each other, therefore orthogonalization needs to be performed in order to use this new basis iteratively. This is done using Gram-Schmidt starting with the smallest $\alpha_i$ and working upwards. The small values of $\alpha$ correspond to diffuse, delocalized orbitals while large values of $\alpha$ correspond to tight, localized orbitals.  Thus, the larger values of $\alpha$ will contain higher momentum components.  The Gram-Schmidt orthogonalization procedure preserves the ordering of the momentum when the input states are sorted by $\alpha_i$. 

We can separate the sum of products and use individual components, $\sum_{m_z,k_z>k_c}s_{m_z,k_z}^{(i)}p_{m_z,k_z}\left(z\right)$ representations as 1D basis functions. For this calculations, $n=6$ and we use the projected STO-$n$G basis functions only for the high momentum part of the Lithium nuclei. The two smallest values of $\alpha_i$ are diffuse and  the part of phase-space these functions describe is described by the Wilson basis. This means that the basis used for the calculation is now
\begin{equation}\label{eq.stohybrid}
\begin{array}{ll}
p_{m_i,k_i},&\quad k<k_c\\
\sum_{m_z,k_z>k_c}s^{(i\p)}_{m_z,k_z}p_{m_z,k_z}\left(z\right),&\quad m=0,k=k_c+i\p,\,i\p>2\\
0,& \quad\mbox{otherwise}
\end{array}
\end{equation}
where $i\p$ labels the Gram-Schmidt orthogonalized states. Applying this basis to the H-Li-H one-electron system with an even symmetrized basis increases the rate of convergence substantially. Using the basis of \Eq{eq.stohybrid} required a basis size of 1493 to obtain the value of -5.1642 $E_h$ while using only the Wilson basis required 4250 basis functions. This comparison was made using $\dx=0.5\sqrt{\pi}$ with 13 Gauss-Jacobi quadarture points and Hermitian matrix vector products.

The use of the STO-$n$G basis' sum-of-products form presented here would most likely be even more useful if one represented the wavefunction as a sum-of-products basis. The reason is that the region of phase-space represented by the STO-$n$G basis only requires the addition of $n$ basis functions. Here, the $n$ basis functions had to be added in each dimension and then coupled into the rest of the basis so the savings was much smaller. That being said, the improvement in basis size is still substantial.  The main downside is that calculating the necessary two-body Coulomb matrix elements for a fixed grid is no longer independent of the nuclear configuration.  This is due to centering of the STO-$n$G basis functions on the nuclei.

\subsection{Other extensions}

There are a few directions one could pursue to increase the efficiency of the Wilson basis: further exploits of the locality, the symmetry, and extending the extrapolation methods to estimate complete basis set limits.

First, it would be beneficial to take advantage of the fact that the individual matrices in the sum-of-products expansion of the Coulomb operator are banded in the sense that the off-diagonal elements decay exponentially. This is somewhat non-trivial to leverage as the bandedness depends on the quadrature point the matrix is evaluated at. In both \Eq{eq.fulloneelec} and \Eq{eq.fulltwoelec}, $p_{\alpha}$ closer to minus one is more diagonal in position space while $v$ closer to one is more diagonal in momentum space. Taking advantage of the bandedness would speed up the calculation and also assist with the troublesome memory usage of $L^4$. 

Second, there room to further exploit the symmetries of the problem.  As mentioned in Section \ref{sec.twoelectron},
the translational symmetry in position space for the two-electron matrix elements has not been exploited here . It may also be useful to use a linear combination of Wilson basis functions that have arguments of the pairs of coordinates  $(x_1,x_2),(y_1,y_2)$ and $(z_1,z_2)$ instead of only $x_i,y_i$ and $z_i$. First, symmetry adaptation of the basis would be more effective since it respects the symmetry of the fully correlated wavefunction. Second, as there is no obvious way to separate the $x_1,x_2$ matrix of \Eq{eq.fulltwoelec}, the use of these basis functions would not decrease the speed of the calculation. 

Third, expanding the basis using the technique of Section \ref{ssec.basis} produced convergence of the calculated energies that can be fitted to a function with form $aM^{b}$. This should make it possible to extrapolate to the complete basis set limit although this was not done in the present study.

The use of phase-space localized (PSL) basis functions has garnered some attention to assist in the calculation of vibrational\cite{Halverson2012,Shimshovitz2012,Brown2016} energy levels. The motivating idea is that only a finite number of PSL basis functions would be needed in order to cover the localized region of phase-space in which the wave function occupies. For vibrational calculations, a comparison with PSL functions and the commonly used Gauss-Hermite functions\cite{Brown2016} or sinc basis\cite{Larsson2016} suggests that this motivating idea has complications. It is important to describe the tail portion of the wavefunction that tunnels outside the classically allowed region of phase-space, but this tail region requires a much larger PSL basis. For vibrational problems, describing the tail region properly is difficult \emph{a priori} as the potential becomes more complicated away from the minimum where a multi-dimensional Taylor expansion is accurate. This is not the case for electronic calculations where the potential can be expanded succinctly away from the nuclei using multipole expansions.  Thus, in the future, it may prove beneficial to replace the tail regions $\left|m\right|>m_c$ with projections of the Wilson basis into standard basis sets.

\section{Conclusion}\label{sec.conc}

Is the Wilson basis competitive with commonly used basis sets at this time? The answer to this question is no. However, there are three advantages worth highlighting. First, this basis performs better in terms of computation and convergence than the previously investigated sinc basis functions.  Second, for isolating a single state, the Wilson basis uses fewer functions than gausslets in the examples tested.  Third, and most promising, we have shown how to combine the Wilson basis functions with the commonly used Slater type Gaussian basis functions.

Future areas of investigation for the Wilson basis include electronic systems with large applied magnetic fields, translationally invariant electronic systems, and use of the Wilson basis in quantum simulation algorithms on quantum computers.  In order to model systems with an applied magnetic field, the extension to complex basis functions must be done and this is relatively straight-forward. Our preliminary computations suggests that the inclusion of a magnetic field does not greatly increase the number of basis functions needed for convergence although $\dx$ must be adjusted.

In this paper, we have only looked at small molecular systems but there is an opportunity to study the electronic properties of solids with the Wilson basis functions.  Since the Wilson basis functions are phase-space localized, the techniques used in this paper can equally be applied to calculations in momentum space.  

A final application area of the Wilson basis function is quantum simulation algorithms for quantum computing.  To utilize the power of quantum computation for electronic structure theory, it is crucial to map fermions to qubits optimally. Efficient mapping allows the number of overall quantum gates applied to be reduced, which is especially important given the current limitations of quantum computing hardware. While there are multiple fermion-to-qubits mappings \cite{Bravyi2000,Seeley2012,Verstraete2005, Ball2005,Bravyi2000,Havlicek2017,Setia2018}, our most recent work has highlighted the potential of the Bravyi-Kitaev Super-Fast mapping \cite{Setia2018} as well as its robustness against certain quantum noise processes \cite{Setia2018a}.  This mapping, which has connections to lattice gauge theory \cite{Zohar2018}, simulates the gauge fields rather than standard mappings that simulate the fermionic fields themselves.  Because of the close connect to lattice gauge theories, our group has noticed strong dependence on the choice of basis set \cite{Hardikar2018}.  Future work will be in applying the Wilson functions described here to fermion encodings that require localized wave functions.

\section*{Acknowledgements}
We gratefully thank the NSF for support under award number 1820747.  

\newpage
\bibliography{citations}
\end{document}